\documentstyle[12pt,epsf]{article}

\def\fnote#1#2{\begingroup\def\thefootnote{#1}\footnote{#2}\addtocounter
{footnote}{-1}\endgroup}

\def\inbar{\vrule height1.5ex width.4pt depth0pt}
\def\IB{\relax{\rm I\kern-.18em B}}
\def\IC{\relax\,\hbox{$\inbar\kern-.3em{\rm C}$}}
\def\ID{\relax{\rm I\kern-.18em D}}
\def\IE{\relax{\rm I\kern-.18em E}}
\def\IF{\relax{\rm I\kern-.18em F}}
\def\IG{\relax\,\hbox{$\inbar\kern-.3em{\rm G}$}}
\def\IH{\relax{\rm I\kern-.18em H}}
\def\II{\relax{\rm I\kern-.18em I}}
\def\IK{\relax{\rm I\kern-.18em K}}
\def\IL{\relax{\rm I\kern-.18em L}}
\def\IM{\relax{\rm I\kern-.18em M}}
\def\IN{\relax{\rm I\kern-.18em N}}
\def\IO{\relax\,\hbox{$\inbar\kern-.3em{\rm O}$}}
\def\IP{\relax{\rm I\kern-.18em P}}
\def\IQ{\relax\,\hbox{$\inbar\kern-.3em{\rm Q}$}}
\def\IR{\relax{\rm I\kern-.18em R}}
\def\IT{\relax{\rm I\kern-.18em T}}
\def\ZZ{\relax{\sf Z\kern-.4em Z}}
\def\a{\alpha}        
 \def\G{\Gamma}    \def\l{\lambda}
     \def\si{\sigma}
\def\Si{\Sigma}   \def\th{\theta}
 
\def\cA{{\cal A}}   \def\cD{{\cal D}}
\def\cF{{\cal F}}   \def\cI{{\cal I}}
   \def\cM{{\cal M}}
\def\cN{{\cal N}}   
\def\cR{{\cal R}}

      \def\tM{\tilde M}   \def\tW{\tilde W}

\def\tchi{\tilde \chi}

 \def\bD{\bar D}
\def\bcD{{\bar \cD}}  
 
  \def\bz{{\bar z}}
   
   \def\bZ{{\bar Z}}
\def\bPhi{{\bar \Phi}} \def\bphi{\bar \phi} 

  \def\bSi{{\bar \Sigma}}
  \def\bth{{\bar \theta}}

\def\fnote#1#2{\begingroup\def\thefootnote{#1}\footnote{#2}\addtocounter
{footnote}{-1}\endgroup}
\def\beq{\begin{equation}}
\def\eeq{\end{equation}}
\def\bea{\begin{eqnarray}}
\def\eea{\end{eqnarray}}
\def\lleq#1{\label{#1}\eeq}
\def\llea#1{\label{#1}\eea}
\let\nn=\nonumber
\def\tabroom{\hbox to0pt{\phantom{\Huge A}\hss}}
\def\notin{\ \hbox{{$\in$}\kern-.51em\hbox{/}}}

\def\del{\partial}

\def\lra{\longrightarrow}

\catcode `\@=11
\def\numberbysection{\@addtoreset{equation}{section}
         \renewcommand{\theequation}{\thesection.\arabic{equation}}}

\begin{document}

\numberbysection

\phantom{ DRAFT--NOT FOR DISTRIBUTION}
\hfill{BONN--TH--94--007}
\vskip 1truein

\centerline{\Large MIRROR SYMMETRY AND STRING VACUA FROM}
\vskip .1truein
\centerline{\Large A SPECIAL CLASS OF FANO VARIETIES}

\vskip .8truein
\centerline{\sc Rolf Schimmrigk
          \fnote{\diamond}{Email address: netah@avzw02.physik.uni-bonn.de}}
\vskip .3truein
\centerline{\it Physikalisches Institut, Universit\"at Bonn}
\vskip .05truein
\centerline{\it Nussallee 12, 53115 Bonn, FRG}

\vskip 1.3truein
\centerline{ABSTRACT}
\vskip .1truein

Because of the existence of rigid Calabi--Yau manifolds, mirror symmetry
cannot be understood as an operation on the space of manifolds with
vanishing first Chern class. In this article I continue to investigate
a particular type of K\"ahler manifolds with positive first Chern
class which generalize Calabi--Yau manifolds in a natural way and which
provide a framework for mirrors of rigid string vacua. This class is
comprised of a special type of Fano manifolds which encode crucial
information about ground states of the superstring. It is shown in
particular that the massless spectra of $(2,2)$--supersymmetric vacua
of central charge $\hat{c}=D_{crit}$ can be derived from special Fano
varieties of complex dimension $(D_{crit}+2(Q-1))$, $Q>1$, and that in
certain circumstances it is even possible to embed Calabi--Yau manifolds
into such higher dimensional spaces.  The constructions described here lead
to new insight into the relation between exactly solvable models and their
mean field theories on the one hand and their corresponding Calabi--Yau
manifolds on the other. It is furthermore shown that Witten's formulation of
the Landau--Ginzburg/Calabi--Yau relation can be applied to the present
framework as well.

\renewcommand\thepage{}
\vfill \eject

\baselineskip=18pt
\parskip .19truein
\parindent=20pt
\pagenumbering{arabic}

{\Large{\bf Contents}}

\begin{enumerate}
\item Introduction
\item A Special Class of Fano Manifolds with Quantized Positive
      1$^{st}$ Chern Class
\item Special Fano Varieties and Critical String Vacua: the Physical Dimension
\item Generalization to Arbitrary Critical Dimension
\item Cohomology of Special Fano Spaces and Critical Spectra
\item When $h^{D_{crit}+Q-1,Q-1} >1$: A Degeneration Phenomenon
\item Splitting and Contraction -- Many Fano Fold Embeddings of Critical Vacua
\item Topological Relations between Fano Varieties
\item Landau--Ginzburg Theories, Special Fano Manifolds, and Critical Vacua
\item Mirrors of Rigid String Vacua
\item Toric Considerations
\item Phases of Special Fano Manifolds
\item Special Complete Intersection Fano Manifolds
\item Conclusion
\end{enumerate}

\vfill \eject

\noindent
\section{Introduction}

\noindent
The discovery of mirror symmetry \cite{cls90}\cite{ls90}\cite{gp90} has made
it possible to address a number of longstanding issues in string theory. Some
of the most important ones are concerned with questions regarding
the correct geometrical
framework of string compactification, the occurrence of topology change and the
appearance of a minimal distance in string theory, the action of modular groups
on moduli space and, finally, the computation of
exact Yukawa couplings not only at special points in the moduli space but as
functions of the moduli.
The latter application is of interest in particular for
concrete application to physics as well as in mathematics: on the physical
side it in fact pre--empties one of the major motivations for considering
string vacua constructed from exactly solvable models whereas on the
mathematical side it allows to obtain results that were thought completely out
of reach prior to the advent of mirror symmetry.

It is not easy, surprisingly, to say, in full generality, what mirror symmetry
is without reference to any particular framework for string compactification.
As a first step one may consider the conventional framework,
formulated in \cite{chsw}, in which left--right symmetric compactifications
without torsion are described by internal spaces which are compact complex
K\"ahler manifolds and admit a covariantly constant spinor,
 i.e. have vanishing first Chern class, so--called Calabi--Yau manifolds.
Such spaces are particularly simple, a  fact that is encoded concisely
in the spectrum of the theory, parametrized, in part, by the cohomology of the
manifolds. Because the internal space is complex, the real cohomology can be
decomposed, via the Hodge decomposition, into complex cohomology groups. Thus
the Betti numbers $b_i={\rm dim}~ H^i(\cM,\IR)$ can be expressed in terms of
the Hodge numbers $h^{(p,q)}={\rm dim}~ H^{p,q}(\cM, \IC)$:
$b_i = \sum_{p+q=i} h^{(p,q)}$.
Since the manifold is K\"ahler, the Hodge numbers are symmetric,
$h^{(p,q)}=h^{(q,p)}$, and because the first Chern class vanishes, it follows
that $h^{(i,0)}=0=h^{(0,i)},~i=1,2$ and $h^{(3,0)}=1=h^{(0,3)}$. Hence the
cohomology of the internal space, summarized in the Hodge diamond

{\footnotesize
\begin{center}
\begin{tabular}{c c c c c c c}
  &  &          &1         &              &   &   \\
  &  &0         &          &0             &   &   \\
  &0 &          &$h^{(1,1)}$ &            &0  &   \\
1 &  &$h^{(2,1)}$ &          &$h^{(2,1)}$ &   &1  \\
  &0 &          &$h^{(1,1)}$ &              &0  &   \\
  &  &0         &          &0             &   &   \\
  &  &          &1         &              &   &   \\
\end{tabular}
\end{center}
}
\noindent
contains only two independent elements $h^{(1,1)}=h^{(2,2)}$ and
$h^{(2,1)}=h^{(1,2)}$ which parametrize the number of antigenerations and
generations, respectively, that we observe in low energy physics.

It has become clear through the work of Gepner
\cite{g88}\cite{g87a}\cite{g87b} on the one hand and Martinec \cite{m89}
and Vafa and Warner \cite{vw89} on the other that an alternative
way of thinking about the same conformal fixed points of the heterotic
string is via the theory ofexactly solvable models and their
mean field theories.
The virtue of the latter formulations lies in the fact that they are
easier to deal with since they are formulated in a
nongeometric fashion. They do, in particular, allow for a detailed
analysis of the massless spectrum
\fnote{1}{In this article attention is focused on the generations and
          antigenerations. The gauge singlets are not considered.}
which naturally falls in the same pattern as the Hodge diamond above
when vacua with central charge nine are considered.

It is in these contexts that mirror symmetry was discovered. In
\cite{cls90} it was observed that for the majority of string vacua
described by Landau--Ginzburg potentials with an isolated singularity
there exists another vacuum for which the number of generations and
antigenerations is reversed, or put differently, the Hodge diamond is
flipped about the diagonal. Since the class of theories constructed in
\cite{cls90} can also be understood as a special class of Calabi--Yau
manifolds this result immediately implies an explicit demonstration of
mirror symmetry as an operation on the space of Calabi--Yau manifolds,
a property of this space that came as quite a surprise to mathematicians.

Even though almost inconceivable, in principle the symmetry found in
\cite{cls90} might have been considered an accidental pairing of a priori
independent theories. That this is not the case was shown in \cite{ls90}
via a construction involving orbifolding and fractional transformations.
It was demonstrated in that paper that explicit mirror maps can indeed easily
be established between many different types of Landau--Ginzburg theories
with mirror spectra by  mapping the orbifold mirror via fractional
transformations of the order parameters onto a mirror Landau--Ginzburg
potential, defining again a complete intersection Calabi--Yau manifold.

In \cite{gp90} an alternative route to mirror symmetry was discovered
in the process of constructing orbifolds of Gepner models. In this framework
mirror symmetry can be described as a Kramers--Wannier type symmetry of the
underlying parafermionic theories.

One of the advantages of the construction of \cite{ls90} is that it allows
to mirror map the complete relevant moduli space \cite{ls94} whereas the
exactly solvable approach is restricted to one particular point in moduli
space,
at which the results of \cite{ls90} of course recover the exactly solvable
mirror theory.

The discovery of mirror symmetry per the constructions of
\cite{cls90}\cite{ls90}\cite{gp90} provided new impulse for the investigation
of string compactification and a number of different avenues have been
addressed vigorously by many groups with two major focal points.
On the one hand the question has been addressed
\cite{kss92,ks94,krsk92,krsk93,bh93,kr93,bh94,rs93} whether mirror symmetry
is an artefact of the particular framework in
which it was discovered, and if not, then what the natural context
is that extends the Landau--Ginzburg/Calabi--Yau framework. On the other hand
the natural question arises what mirror symmetry can teach us about string
theory. In a nutshell, mirror symmetry provides a tool that allows access
to problems that before its discovery in the form of \cite{cls90,ls90,gp90}
were impossible to solve. The main
foci in this direction are issues regarding the action of the modular group on
moduli space \cite{cogp91,df93,cdr93,c93}, questions concerning a mild type
of topology change \cite{agm93}, and the computation of Yukawa couplings,
pioneered in \cite{cogp91}. The latter has led to a breakthrough in
mathematics and mirror symmetry is presently being used in this context
as a hypothesis which leads to novel results in algebraic geometry. Whenever
independent checks are possible, mirror symmetry has been shown to lead to
correct results in all computations that have been performed
so far for genus zero instantons
\cite{cogp91,norwegians,m92,f93,kt93,kt93b,lt93,bs93,cofkm93,hkty93,
 cfkm94,bdhj93}
as well as higher genus instantons \cite{bcov93I, bcov93II} in Calabi--Yau
manifolds of complex dimension three, and for genus zero instantons in
Brieskorn--Pham type  Calabi--Yau manifolds for
$dim_{\IC} M > 3$ \cite{gmp94} \cite{ka93}.

The present paper is concerned with the first question -- what is the
proper framework for mirror symmetry? As far as sheer numbers are
concerned it appears that the space of Calabi--Yau manifolds is the
structure to go for because, very roughly, there are more Landau--Ginzburg
models known than there are exactly solvable models and it is clear
that many more Calabi--Yau manifolds can be constructed than there
are Landau--Ginzburg theories. However, the space of all Calabi--Yau
manifolds has a disconcerting property:  it contains spaces which are
rigid, i.e. they do
not have string modes corresponding to complex deformations of the
manifold, fields that describe generations in the low energy
theory. Since mirror symmetry exchanges complex deformations and K\"ahler
deformations of a manifold, the latter describing the antigenerations
seen by a four--dimensional observer, it would seem that the mirror of a
rigid Calabi--Yau manifold cannot be K\"ahler and hence does not describe a
consistent string vacuum. In fact, it appears that the mirror vacuum cannot
even be $N=1$ spacetime supersymmetric \cite{z79}.
It follows then that the class of Calabi--Yau manifolds is not the
appropriate setting by a long shot in which to discuss mirror symmetry and
the question arises  what the proper framework might be.

In this article I continue an investigation, initiated in \cite{rs93}, of a
class of manifolds which generalizes the class of Calabi--Yau spaces of complex
dimension $D_{crit}$ in a natural way. The manifolds involved are of complex
dimension $(D_{crit}+2(Q-1))$, $Q>1$, and feature a positive first Chern class
that is quantized in a particular way. Thus they do not
describe, a priori, consistent string groundstates. Surprisingly however, it is
possible to derive from these higher dimensional manifolds the spectrum of
critical string vacua. This can be done not only for the generations but also
for the antigenerations.  Moreover a construction exists which for particular
types of these new manifolds allows the derivation of the corresponding
$D_{crit}$--dimensional Calabi--Yau manifold directly from the
$(D_{crit}+2(Q-1))$--dimensional space.

This new  class of manifolds is, however, not in one to one correspondence
with the class of Calabi--Yau manifolds as it contains manifolds which
describe string vacua that do not contain massless modes corresponding to
antigenerations.  It is precisely this new type of manifold that is
needed in order to construct mirrors of rigid Calabi--Yau manifolds
without generations.

Hence the class of special Fano manifolds appears as a natural generalization
of the standard Calabi--Yau framework. Further evidence that important
Calabi--Yau constructions admit generalizations to special Fano manifolds
has been presented in two recent papers \cite{cdp93,bb94}.
 It has been shown in \cite{cdp93} that the computation of Yukawa couplings
along the lines of \cite{cogp91} can be applied in this framework as well,
whereas  Batyrev and Borisov \cite{bb94} present an interesting formulation
of our construction in the framework of toric geometry.
These authors generalize Batyrev's  Calabi--Yau mirror construction
\cite{vitja92} and do find, as conjectured in \cite{rs93}, a combinatorical
understanding of mirror manifolds of Calabi--Yau manifolds of codimension
larger than one by utilizing special Fano manifolds.

The paper is organized as follows: in Section 2 the manifolds in
question will be briefly reviewed and their salient defining
properties described. In Section 3 it will be shown how the spectrum of
critical string vacua can be derived from the special class of Fano
manifolds of Section 2. It will also be shown that for a
number of different classes it is possible to derive the critical
$D_{crit}$--dimensional Calabi--Yau manifold directly from higher dimensional
Fano manifolds. In Section 4 the geometrical construction of Section 3 will
be generalized to arbitrary critical dimensions.
Section 5 contains a more systematic discussion of the higher dimensional
cohomology and Section 6 is devoted to an important `degeneration
phenomenon'. Sections 7 and 8 contain a discussion of the consequences
for the present class of manifolds of the splitting and contraction
construction of \cite{cdls88} as well as the topological identities that
were used in \cite{cdls88} for the analysis of the
class of complete intersection Calabi--Yau manifolds.
In Section 9 it is shown how the geometrical construction of Section
3 pertains to the standard discussion of the Landau--Ginzburg/Calabi--Yau
connection and Section 10 is devoted to the discussion of mirrors of
rigid Calabi--Yau spaces. Two final Sections contain brief remarks
regarding toric considerations and the generalization of the
hypersurfaces of Section 2 to weighted complete intersections.

\vskip .2truein
\noindent
\section{A Special Class of Fano Manifolds with Quantized Positive
         First Chern Class }
\vglue 0.4cm
\noindent
The construction of  noncritical manifolds proceeds via the following
prescription \cite{rs93}:
\begin{itemize}
\item Fix the central charge $c$ of the (2,2)--vacuum states
     and its critical dimension
 \beq
  D_{crit} = c/3.
 \eeq
\item Choose a positive integer $Q\in \IN$.
\item Introduce $(D_{crit}+2Q)$ complex coordinates
     $(z_1,...,z_{D_{crit}+2Q}), z_i \in \IC$.
\item Define an equivalence relation
   \beq
     (z_1,...,z_{D_{crit}+2Q})
     \sim (\l^{k_1}z_1,...,\l^{k_{D_{crit}+2Q}} z_{D_{crit}+2Q})
   \eeq
 where $\l \in \IC^*$ is a nonzero complex number and the
positive integers  $k_i \in \IN$ are the weights of these coordinates.
The set of these
equivalence classes defines so--called weighted projective spaces,
compact manifolds which will be denoted by $\IP_{(k_1,...,k_{D_{crit}+2Q})}$.
\item Define hypersurfaces in the ambient weighted projective space
    by imposing a constraint defined by polynomials $p$ of degree
  \beq
   d= \frac{1}{Q} \sum_i k_i
  \lleq{pquant}
  i.e. $p(\l^i z_i) = \l^d p(z_i)$.
\end{itemize}

\noindent
The family of hypersurfaces embedded in the ambient space
 as the zero locus of $p$ will be denoted by
\bea
 M_{D_{crit}+2(Q-1)} &=&  \{p(z_1,\dots,z_{D_{crit}+2Q})=0\}~\cap ~
                          \IP_{(k_1,\dots, k_{D_{crit}+2Q})} \nn \\
                   &=&  \IP_{(k_1,...,k_{D_{crit}+2Q})}
                      \left[ \frac{1}{Q} \sum_{i=1}^{D_{crit}+2Q} k_i\right]
\llea{spefam}
and will be called a configuration.

Relation (\ref{pquant}) is the defining
property of the class of spaces to be considered below. It is a rather
restrictive condition in that it excludes many types of varieties which
are transverse and even smooth but are not of physical relevance
\fnote{2}{It will become clear below that this definition is rather
          natural in the context of the theory of Landau--Ginzburg
          string vacua with an arbitrary number of scaling fields. A
         particular simple manifold in this class, the cubic sevenfold
       $\IP_8[3]$, has been the subject of refs.
       \cite{cdp93}\cite{v92}.}.
A simple example is the Fermat type hypersurface
\beq
\IP_{(420,280,210,168,140,120,105)}[840] \ni
\{p= \sum_{i=1}^7 z_i^{i+1} =0 \}
\lleq{excluded}
which is not only quasihomogeneous but also transverse, i.e. a
quasismooth manifold. The fact that this quasismooth variety is rather
different from the manifolds of type (\ref{spefam}) can be understood in
a more geometrical way. Namely, what makes the Fano varieties of type
(\ref{spefam}) special is, among other things, a property that they share
with Calabi--Yau manifolds: they do not contain singular sets of codimension
one. For Calabi--Yau spaces this means that
the singular sets are either points or (complex) curves, but never surfaces.
In a more general context this fact translates into the statement that the
only resolutions that have be performed are so--called small resolutions.
The above Fermat type hypersurface (\ref{excluded}) on the other hand
contains the singular codimension--1 four--fold
$S=\IP_{(210,140,105,84,70,60)}[420]$.

Alternatively, Fano manifolds of the special type (\ref{spefam})
above may be characterized
via a curvature constraint. Because of (\ref{pquant}) the first
Chern class is given by
\beq
c_1(M_{D_{crit}+2(Q-1)}) =(Q-1)~c_1(\cN)
\lleq{c1quant}
where $c_1(\cN)=dh$ is the first Chern class of the normal bundle
$\cN$ of the hypersurface $M_{D_{crit}+2(Q-1)}$ and $h$ is the pullback of
the K\"ahler form
${\rm H}\in {\rm  H}^{(1,1)}\left(\IP_{(k_1,\dots,k_{D_{crit}+2Q})}\right)$
of the ambient space. Hence the first Chern class is quantized in multiples
of the degree of the hypersurface $M_{D_{crit}+2(Q-1)}$.
For $Q=1$ the first Chern class vanishes and the manifolds
for which (\ref{pquant}) holds are Calabi--Yau manifolds, defining
consistent groundstates of the supersymmetric closed string.
For $Q > 1$ the first Chern class is nonvanishing and therefore these
manifolds cannot possibly describe vacua of the critical string, or
so it seems.

 It has been shown in \cite{rs93} however that it is possible to
derive from these higher dimensional manifolds the massless spectrum
of critical vacua. It is furthermore possible, for certain subclasses
of hypersurfaces of type (\ref{spefam}), to construct
Calabi--Yau manifolds $M_{CY}$ of dimension $D_{crit}$  and complex
codimension
\beq
codim_{\IC} (M_{CY}) =Q      \nn
\eeq
directly from these manifolds.
The integer $Q$ thus plays a central r\^{o}le:
the critical dimension is the dimension of the noncritical manifolds offset
by twice the coefficient of the
first Chern class of the normal bundle of the hypersurface, which
involves $Q$. The physical interpretation of the integer $Q$ is that
of a total charge associated to the corresponding Landau--Ginzburg
theory which determines the codimension of the Calabi--Yau manifold
which it describes.

As mentioned already in the introduction the class of spaces defined by
(\ref{spefam}) contains manifolds with no antigenerations
and hence it is necessary to have some way other than Calabi--Yau
manifolds to represent string groundstates if one intends to understand
the relation of the latter to higher dimensional special Fano varieties.
One possible way to do this is to relate them to Landau--Ginzburg theories:
any manifold of type (\ref{spefam}) can be viewed
as a projectivization of an affine noncompact hypersurface
\beq
\IC_{(k_1,...,k_{N+2})}\left[d\right] \ni \{p(z_1,...,z_{N+2})=0\}
\lleq{affvar}
via a weighted equivalence relation
\beq
(z_1,\dots ,z_{N+2}) \equiv (\l^{k_1}z_1,\dots, \l^{k_{N+2}}z_{N+2}).
\eeq
Because the polynomial $p$ is assumed to be transverse in the projective
ambient
space the affine variety has a very mild singularity: it has an isolated
singularity at the origin defining what is called a catastrophe in the
mathematical literature. The polynomial ring of such a catastrophe is
isomorphic to the chiral ring of a N$=$2 supersymmetric
Landau--Ginzburg theory and hence allows us to compute physical quantities
via methods in this framework
\fnote{3}{Landau--Ginzburg theories are discussed in greater detail in
          a different context in Section 9.}.

Even though the manifolds (\ref{spefam}) therefore correspond to
LG theories of central charge $c=3D_{crit}$ they can, however,
not be identical to such theories: Consider the
case when the critical dimension of the internal space corresponds
to our world, i.e. $D_{crit}=3$ and $Q=2$. The cohomology
of $M_5$ leads to the Hodge diamond

\begin{scriptsize}
\begin{center}
\begin{tabular}{c c c c c c c c c c c}
  &  &      &   &          &1         &         &   &       &   &  \\
  &  &      &   &0         &          &0        &   &       &   &  \\
  &  &      &0  &          &$(1,1)$     &         &0  &       &   &  \\
  &  &0     &   &0         &          &0        &   &0      &   &  \\
  &0 &      &0  &          &$(2,2)$     &         &0  &       &0  &  \\
0 &  &$(4,1)$ & &$(3,2)$   &          &$(3,2)$    &   &$(4,1)$  &   &0 \\
  &0 &      &0  &          &$(2,2)$     &         &0  &       &0  &  \\
  &  &0     &   &0         &          &0        &   &0      &   &  \\
  &  &      &0  &          &$(1,1)$     &         &0  &       &   &  \\
  &  &      &   &0         &          &0        &   &       &   &  \\
  &  &      &   &          &1         &         &   &       &   &  \\
\end{tabular}
\end{center}
\end{scriptsize}
where $(p,q)$ denotes the dimension $h^{(p,q)}$ of the cohomology group
${\rm H}^{(p,q)}(M_5)$.

It is clear from this Hodge diamond that the higher dimensional manifolds
will contain more modes than the critical vacuum and hence the relation
of the spectrum of the critical vacuum and the cohomology of the noncritical
manifolds will be a nontrivial one. It is this question which is the subject
of the next Section.

\vskip .2truein
\noindent
\section{Special Fano Varieties and Critical String Vacua:
        the Physical Dimension}

\noindent
Even though for $Q>1$ the spaces (\ref{spefam}) clearly are not  of
Calabi--Yau type and hence do not, a priori, lead to consistent string
vacua, they do in fact encode essential information about these string vacua.
Most importantly it is possible to
derive from these higher dimensional manifolds the massless spectrum
of critical vacua. Furthermore we will see that for certain subclasses
of hypersurfaces of type (\ref{spefam}) it is possible to construct
Calabi--Yau manifolds $M_{CY}$ of dimension $D_{crit}$  and complex
codimension $codim_{\IC} (M_{CY}) =Q$
directly from these manifolds.

 A first step in the analysis of any physical theory is to determine
its physical spectrum; hence the first question to address is how it is
possible to derive the massless spectrum of the corresponding string vacua
from these higher dimensional manifolds. In order to be able to formulate
our geometrical framework in a concise way it is useful to recall some of
the salient characteristics of hypersurfaces
in weighted projective spaces. The feature most relevant for the following
considerations comes from the fact that weighted projective spaces are
singular objects, the singularities being of orbifold type. Submanifolds
defined via some polynomial constraints will, in general, intersect these
singularities and thus be singular themselves. Such singular spaces,
called V--manifolds in older references, so far have not been of interest
for string physics, rather it is the smoothed out, `resolved',  manifold
which has been the physically important object. Such is the
case for the critical situation $Q=1$.

For $Q>1$ it is clear that a projection is needed which selects from the
cohomology of the Fano variety the physically relevant part. In the
following it will be shown that the following selection rule provides
for a geometrical projection of
the Hodge diamond of the special Fano manifold.

\begin{quote}
  {\it The generation and antigeneration spectrum of a c$=$$D_{crit}$
          vacuum of the N=1 spacetime supersymmetric string is parametrized
          by the cohomology carried by the singular subvarieties of the
      Fano fold together with the monomials of the integrally charged subring
          of the polynomial ring.}
\end{quote}

It is important to realize that our prescription is radically different
from the
$Q=1$ case where not only the cohomology carried by the singular sets is
important but also the cohomology induced by the fibers of the bundles
that are introduced in the process of resolving these singularities.

Of immediate concern, of course, are vacua which are mirrors of rigid
Calabi--Yau manifolds: Consider the orbifold $T_1^3/\ZZ_3^2$ where the
two actions are defined as
$(z_1,z_4) \longrightarrow  (\a z_1,\a^2 z_4)$, all other coordinates
invariant and $(z_1,z_7) \longrightarrow (\a z_1,\a^2 z_7) $, all other
invariant.  Here $\a$ is the third root of unity.
The resolution of the singular orbifold leads to a Calabi--Yau manifold
with 84 antigenerations and no generations \cite{gvw89}.
This is precisely the mirror flipped spectrum of the exactly solvable
tensor model $1^9$ of 9 copies of $N=2$ superconformal minimal models at
level $k=1$ \cite{g88} which can be described in terms of the
Landau--Ginzburg potential $W=\sum z_i^3$ which belongs to the configuration
$\IC_{(1,1,1,1,1,1,1,1,1)}[3]$. The generations of this theory are obtained
by counting the number of monomials of charge unity, of which there are 84.
Imposing the GSO projection by modding out a $\ZZ_3$ symmetry this
Landau--Ginzburg theory does not lead to new generations and furthermore
does not create antigenerations. Thus this LG theory leads to the
same spectrum as the $1^9$ model
\fnote{3}{ It would seem that an obvious generalization of
       this 7--dimensional smooth manifold is the infinite class of models
           $\IC_{(1,1,1,1,1,1,1,1,1+3p)}[3+p]$, but since the
           manifolds (\ref{spefam}) are required to be transverse the only
           possibility is $p=0$.}.

This Landau--Ginzburg theory clearly is a mirror candidate for the
resolved torus orbifold just mentioned \cite{cdp93}\cite{v92}
and the question arises whether a manifold corresponding to
this LG potential can be found. Since the theory does not contain modes
corresponding to (1,1)--forms it appears that the manifold cannot be
K\"ahler and hence not projective. Thus it appears that the
7--dimensional manifold $\IP_8[3]$ whose polynomial ring is identical to
the chiral ring of the LG theory is merely useful as an auxiliary device
in order to describe one sector of the critical LG string vacuum.

There exists one further theory in the space of all Landau--Ginzburg vacua
which does not have any antigenerations and which is distinct from the one
discussed above. Like the previous example it originates from an exactly
solvable model in the class of Gepner models
\beq
(2^6)_{A^6}^{(0,90)} \equiv \IC^*_{(1,1,1,1,1,1,2)}[4] \ni
\left\{\sum_i z_i^4 + z_7^2 =0\right\}
\eeq
and leads to an obviously smooth manifold $\IP_{(1,1,1,1,1,1,2)}[4]$ as
well.

Our geometrical projection immediately explains why mirrors of rigid vacua are
so different: in the context of special Fano manifolds they are
distinguished by the fact that they do not lead to K\"ahler modes because
they are described by smooth spaces. Thus these two theories fit in
very well in our way of thinking
about the critical spectrum of the corresponding string vacuum.

Vacua without antigenerations are rather exceptional however; the generic
groundstate will have both sectors, generations and antigenerations.
The idea described above to derive the spectrum works for
higher dimensional manifolds corresponding to different types of critical
vacua, as will be shown now.

To be concrete consider the
exactly solvable tensor theory $(1\cdot 16^3)_{A_2 \otimes E^3_7}$ with
35 generations and 8 antigenerations which
corresponds to a Landau--Ginzburg theory belonging to the configuration
\beq
\IC^*_{(2,3,2,3,2,3,3)}[9]^{(8,35)}
\eeq
and which induces, via projectivization, a 5--dimensional weighted
hypersurface
\beq
\IP_{(2,2,2,3,3,3,3)}[9] \ni
\left\{p=\sum_{i=1}^3 (y_i^3x_i+x_i^3)+x_4^3=0\right\},
\lleq{ex3}
which contains the orbifold singularity sets
\bea
\ZZ_3 &:& \IP_3[3] \ni \left\{p_1=\sum_{i=1}^4 x_i^3=0\right\},  \nn \\
\ZZ_2 &:& \IP_2.
\llea{sings}
The $\ZZ_3$--singular set is a smooth cubic surface which supports
 seven (1,1)--forms which follows from $\chi(\IP_3[3])=9$ and the fact that
$c_1(\IP_3[3])>0$. The $\ZZ_2$ singular set is just the projective plane and
therefore adds one further (1,1)--form.  Hence the singularities induced on
the 5--fold by the singularities of the ambient weighted projective space
$\IP_{(2,2,2,3,3,3,3)}$ give rise to a total of eight  (1,1)--forms. A simple
count leads to the result that the subring of monomials of charge unity is of
dimension 35.  Thus we have derived the
spectrum of the critical theory from the noncritical manifold
(\ref{ex3}).

The singularity structure (\ref{sings}) of the manifold under consideration is
rather intriguing and suggests that the idea introduced above of relating the
spectrum of the string vacuum to the singularity structure of the Fano variety
implies much more:  namely it leads also to a prescription of how
 to derive from these higher dimensional
manifolds the Calabi--Yau manifold of critical dimension! Thus a
canonical prescription is obtained which also allows to pass from the
Landau--Ginzburg theory to its geometrical counterpart.

This works as follows:
 Recall that the structure of the singularities of the weighted
hypersurface only involved part of the superpotential, namely the cubic
polynomial $p_1$ which determined the $\ZZ_3$ singular set described by a
surface.  The superpotential thus splits naturally into the two parts
\beq
p=p_1 + p_2
\eeq
where $p_2$ is the remaining part of the polynomial. The idea now is to
consider the product $\IP_3[3] \times \IP_2$, where the factors are
determined by the singular sets of the higher dimensional space, and to
impose on this 4--dimensional space a constraint described by  the remaining
part of the polynomial which did not take part in the definition of the
singularities of the special Fano 5--fold.  In the case at hand this
leaves a polynomial of bidegree $(3,1)$ and hence we are lead to a manifold
embedded in the configuration
\beq
\matrix{\IP_2 \hfill \cr \IP_3\cr}
\left[\matrix{3&0\cr 1&3\cr}\right]
\lleq{ex3mine}
via the  polynomials
\bea
p_1 &=& y_1^3x_1 + y_2^3x_2 +y_3^3x_3~, \nn \\
p_2 &=& \sum_{i=1}^4 x_i^3.
\eea
This  is precisely the Calabi--Yau 3--fold constructed in \cite{rs87},
the exactly solvable model of which was later found in \cite{g87b}. Thus we
have found how to construct from the noncritical manifold (\ref{ex3})
a critical Calabi--Yau manifold of the correct physical dimension.

A class of manifolds of a different type which can be discussed in
this framework rather naturally is defined by
\beq
\IP_{(2k,K-k,2k,K-k,2k_3,2k_4,2k_5)}[2K]
\lleq{niceclass}
where $K=k+k_3+k_4+k_5$. Assume, for the moment, that  $K/k$ and $K/k_i$
are integers. The  potentials can then be chosen to be of the form
\beq
W=\sum_{i=1}^2 (x_i^{K/k}+x_iy_i^2)+x_3^{K/k_3}+x_4^{K/k_4}+x_5^{K/k_5}.
\eeq
and the singularities of these manifolds are
\bea
\ZZ_2 &:&~~\IP_{(k,k,k_3,k_4,k_5)}[K]~\ni~
          \left\{\sum_i x_i^{K/k_i}=0\right\}~, \nn \\
\ZZ_{K-k} &:&~~ \IP_1.
\eea

This suggests that there exists a canonical way to proceed to the
corresponding Calabi--Yau manifold of these theories by considering
manifolds embedded in
\beq
\matrix{\IP_1 \hfill \cr \IP_{(k,k,k_3,k_4,k_5)}\cr}
\left[\matrix{2&0\cr k&K\cr}\right],
\lleq{myclass}
defined by polynomials
\bea
p_1 &=& y_1^2x_1 + y_2^2x_2 \nn \\
p_2 &=& x_1^{K/k}+ x_2^{K/k} + x_3^{K/k_3} + x_4^{K/k_4} + x_5^{K/k_5}.
\eea
That this correspondence is in fact correct, can be inferred from the
work of \cite{rs89} where it was shown that these codimension--2 weighted
complete intersection Calabi--Yau manifolds correspond
to $N=2$ minimal exactly solvable tensor models  of the type
\beq
\left[ 2\left(\frac{K}{k}-1\right)\right]_D^2\otimes
\prod_{i=3}^5  \left(\frac{K}{k_i}-2\right)_{A}.
\eeq
where the subscripts indicate the type of the affine invariants chosen for
the individual levels.

The constraints that $k$ and the $k_i$ divide $K$ have been imposed
purely for convenience since it allows for a nice coherent description
of the resulting class.
These restrictions however can be relaxed to more general
types of potentials, the discussion of which will take up the remainder
of this Section. The types include, among others:

\noindent
$\bullet$ {\bf Single tadpole} involving, say, the variables $x_4,x_5$
\beq
x_4^a + x_4x_5^b
\eeq
which leads to the modified divisibility criterion
$\frac{K}{k_4},\frac{1}{k_5}(K-k_4)\in \IN$. An example is furnished by
\beq
\IP_{(7,2,7,2,4,4,6)}[16]\ni
\left\{p=\sum_{i=1}^2 \left(x_i^2y_i + y_i^8\right)
             +y_3^4+y_4^4+y_4y_5^2=0\right\}
\eeq
which leads to
\beq
\matrix{\IP_1 \hfill \cr \IP_{(1,1,2,2,3)}\cr}
\left[\matrix{2&0\cr 1&8\cr}\right]
\ni
\left\{ \begin{array}{c l}
        p_1=& \sum_{i=1}^2 x_i^2y_i=0 \\
        p_2=& y_1^8+y_2^8+y_3^4+y_4^4+y_4y_5^2 =0 \\
         \end{array}
\right\}.
\eeq

\noindent
$\bullet$ {\bf Isolated double tadpole}
\beq
x_3^a+x_3x_4^b+x_4x_5^c
\eeq
involving three coordinates for which
$\frac{K}{k_3}, \frac{1}{k_4}(K-k_3), \frac{1}{k_5}(K-k_4)\in \IN$.
An example of this type is obtained via
\beq
\IP_{(17,2,17,2,8,12,14)}[36]\ni
\left\{p=\sum_{i=1}^2 \left(x_i^2y_i + y_i^{18}\right)
               +y_3^3y_4+y_4^3+y_3y_5^2=0\right\}
\eeq
from which our construction derives the Calabi--Yau 3--fold
\beq
\matrix{\IP_1 \hfill \cr \IP_{(1,1,4,6,7)}\cr}
\left[\matrix{2&0\cr 1&18\cr}\right]
\ni
\left\{ \begin{array}{c l}
        p_1=& \sum_{i=1}^2 x_i^2y_i=0 \\
        p_2=& y_1^{18}+y_2^{18}+y_3^3y_4+y_4^3+y_3y_5^2=0 \\
         \end{array}
\right\}.
\eeq

\noindent
$\bullet$ {\bf Nonisolated tadpoles.}
It is not necessary however to have the variables $x_3,x_4,x_5$ completely
decoupled from the rest of the coordinates. Examples abound where
they are linked nontrivial ways to other variables as well.
This is illustrated by the 5--fold
\beq
\IP_{(3,4,2,3,4,2,2)}[10]\ni
\left\{p=\sum_{i=1}^2 \left(x_i^2y_i + y_i^2 z_i+z_i^5\right)
                   + z_3^5=0\right\}
\eeq
from which we can derive the Calabi--Yau 3--fold
\beq
\matrix{\IP_1 \hfill \cr \IP_{(2,2,1,1,1)}\cr}
\left[\matrix{2&0\cr 2&5\cr}\right]
\ni
\left\{ \begin{array}{c l}
        p_1=& \sum_{i=1}^2 x_i^2y_i=0 \\
        p_2=& \sum_{i=1}^2 (y_i^2z_i+z_i^5)+z_3^5 =0 \\
         \end{array}
\right\}.
\eeq

\noindent
$\bullet$ {\bf Exceptional types.}
Finally, it should be emphasized that our construction also works for
polynomials which contain so--called exceptional terms, an example in
point being
\beq
\IP_{(2,4,5,2,4,5,6)}[14]\ni
\left\{p=\sum_{i=1}^2 \left(x_i^2y_i + y_i^3 z_i+z_i^7\right)
                      + z_1u+uy_1^2=0\right\}
\eeq
leading to the Calabi--Yau 3--fold
\beq
\matrix{\IP_1 \hfill \cr \IP_{(2,2,1,1,3)}\cr}
\left[\matrix{2&0\cr 2&7\cr}\right]
\ni
\left\{ \begin{array}{c l}
        p_1=& \sum_{i=1}^2 x_i^2y_i=0 \\
        p_2=& \sum_{i=1}^2 (y_i^3z_i+z_i^7)+z_1u+uy_1^2 =0 \\
         \end{array}
\right\}.
\eeq

It is clear from this discussion that the spaces described by
(\ref{myclass}) describe a much wider class of models than the known
exactly solvable spaces which were the focus of ref. \cite{rs89}.

The picture that emerges from these constructions then is the
following: embedded in the higher dimensional manifold is a submanifold
which is fibered.  The base and the fibers of this fibration are determined
by the singular sets of the ambient manifold. The Calabi--Yau manifold
itself is a hypersurface embedded in this fibered  submanifold. A heuristic
sketch of the geometry is shown in the Figure 1.

\vskip .2truein
\centerline{ \epsfbox{figure1.eps} }

\noindent
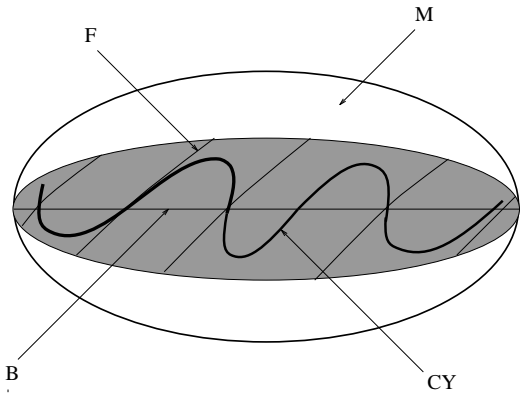
{\bf Figure 1:} {\it The Calabi-Yau manifold CY is embedded in the fibered
        submanifold $B\times F$ of the higher dimensional manifold $M$. Both
     the base $B$ and the fibers $F$ of the bundle (indicated by the shaded
    surface) are determined by the singular sets of the manifold $M$.}

The examples above illustrate the simplest situation that can appear.
In more complicated manifolds the singularity structure will consist
of hypersurfaces whose fibers and/or base themselves are fibered,
leading to an iterative procedure. Figure 2 depicts more general situations
of such a type.

\vskip .2truein
\centerline{ \epsfbox{figure2.eps} }

\noindent
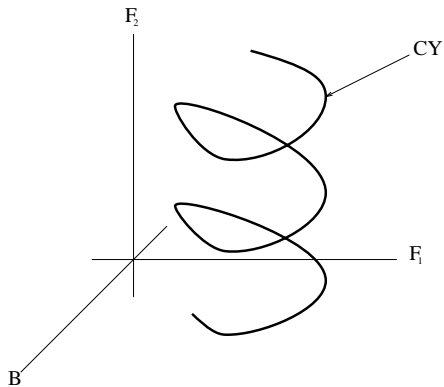
{\bf Figure 2:} {\it The Calabi-Yau manifold CY is embedded in the fibered
    submanifold $B\times F_1 \times \cdots \times F_n$ determined by the
    singular sets of the  higher dimensional manifold $M$.}

 The fibered submanifold of the higher dimensional Fano variety
will, in those cases, be of codimension larger than one and the
Calabi--Yau manifold will be described by a submanifold with
codimension larger than one as well. To illustrate this point consider
the 9--fold
\beq
\IP_{(5,5,6,6,6,4,4,4,8,8,8)}[16] \ni
\{ \sum_{i=1}^2 \left(u_i^2v_i+v_i^2w_i+w_i^2x_i +x_i^2\right)
   + v_3^2w_3+w_3^2x_3 +x_3^2 =0\}.
\eeq
The $\ZZ_2$--fibering leads to the split
\beq
\matrix{\IP_1\hfill \cr \IP_{(3,3,3,2,2,2,4,4,4)}\cr}
\left[\matrix{2 &0 \cr
              3 &16 \cr}\right]
\eeq
which in turn leads to the $\ZZ_2$ split
\beq
\matrix{\IP_1\hfill \cr \IP_2 \hfill\cr \IP_{(1,1,1,2,2,2)}\cr}
\left[\matrix{2 &0 &0 \cr
              1 &2 &0 \cr
              0 &1 &4 \cr}\right]
\eeq
which finally leads to
\beq
\matrix{\IP_1 \cr \IP_2\cr \IP_2 \cr \IP_2\cr}
\left[\matrix{2 &0 &0 &0\cr
              1 &2 &0 &0\cr
              0 &1 &2 &0\cr
              0 &0 &1 &2\cr}\right]  \ni
\left\{ \begin{array}{c l}
p_1 &= \sum u_i^2v_i \nn \\
p_2 &= \sum v_i^2w_i \nn \\
p_3 &= \sum w_i^2x_i \nn \\
p_4 &= \sum x_i^2
         \end{array}
\right\}.
\lleq{codim4}
Thus the 9--fold fibers iteratively and the splits of the polynomial
$p$ are dictated by the fibering.

A further manifold leading to a critical vacuum with codimension larger than
two and which is of a type that will make another appearence later in Section 9
is  the 7--fold
\beq
\IP_{(1,1,6,6,2,2,2,2,2)}[8]~\ni~
\left\{\sum_{i=1}^2 \left(x_i^2y_i + y_iz_i +z_i^4\right)
 +z_3^4+z_4^4+z_5^4=0\right\}
\eeq
which leads to the $\ZZ_2$ fibering
$\IP_1\times \IP_{(3,3,1,1,1,1,1)}[4]$ which in turn leads to the
$\ZZ_3$ fibering $\IP_1\times \IP_1 \times \IP_4[4]$
and finally to the Calabi--Yau configuration
\beq
\matrix{\IP_1\cr \IP_1 \cr \IP_4\cr}
\left[\matrix{2&0&0\cr
              1&1&0\cr
              0&1&4\cr}\right] \ni
\left\{ \begin{array}{c l}
        p_1=& \sum_{i=1}^2 x_i^2y_i=0 \\
        p_2=& \sum_{i=1}^2 y_iz_i =0 \\
        p_3=& \sum_{j=1}^5 z_i^4 =0
         \end{array}
\right\}
\lleq{codim3.1}
which is of codimension 3.
This example is of interest because it shows, as will become clear later,
that there are
nontrivial relations between the higher dimensional special Fano manifolds.
This will follow from the processes of splitting and contraction
of Calabi--Yau manifolds introduced in ref. \cite{cdls88} .
It will become obvious that in fact the Calabi--Yau manifold
(\ref{codim3.1})
is an ineffective split of a Calabi--Yau manifold in the class
(\ref{niceclass}). Thus there also exists a corresponding
relation between the higher dimensional manifolds.

\vskip .2truein
\noindent
\section{Generalization to Arbitrary Critical Dimensions}

\noindent
Even though the examples discussed in the previous Section were all
concerned with 6--dimensional Calabi--Yau manifolds and the way they are
embedded in the new class of special Fano spaces, the ideas introduced in
\cite{rs93} and described in greater detail above are not specific to
this dimension.  Instead of considering compactifications down to the
physical dimension, namely four, one might contemplate compactifying
down to two, six or eight dimensions, or else, discuss the class of manifolds
(\ref{spefam}) not in the context of compactification at all.

To illustrate this point consider the doubly infinite class of
$(m+n)$--dimensional complex compact manifolds
\beq
\IP_{(m+1,n-1,m+1,n-1,\dots ,m+1,n-1,m+1,\cdots ,m+1)}[(m+1)n]
\eeq
with $(m+1)$ pairs of coordinates with weights $(m+1,n-1)$  and $(n-m)$
coordinates of weight $(m+1)$, defined by polynomials
\beq
p=\sum_{i=1}^{m+1} (x_i^n+x_iy_i^{m+1}) + \sum_{j=m+2}^{n+1} x_j^n.
\eeq
These $(m+n)$--dimensional spaces lead to $Q=2$ and hence define, according
to our construction, Calabi--Yau manifolds of dimension $(m+n-2)$
embedded in the configurations
\beq
\matrix{\IP_m\cr \IP_n\cr}\left[\matrix{(m+1)&0\cr 1&n\cr}\right]
\lleq{infser}
via the zero locus of the polynomials
\beq
p_1 = \sum_i y_i^{m+1} x_i,~~~~~~~~p_2 = \sum_{i=1}^{n+1} x_i^n.
\eeq

The simplest example is, of course, the case $n=2$ where the higher
dimensional manifold is a 3--fold described by
\beq
\IP_{(2,1,2,1,2)}[4]~\ni ~
\left\{\sum_{i=1}^2 (z_i^2+z_iy_i^2)+z_3^2=0\right\}
\eeq
with a $\ZZ_2$ singular set isomorphic to the sphere
$\IP_2[2]\sim \IP_1$ which contributes one (1,1)--form.
The singularity structure of the 3--fold then relates this space to the
complex torus described by the algebraic curve
\beq
\matrix{\IP_1 \cr  \IP_2}\left[\matrix{2 &0\cr  1 & 2\cr}\right].
\eeq
The Landau--Ginzburg theory corresponding to this
theory derives from an exactly solvable tensor model $(2^2)_{D^2}$
described by two $N=2$ superconformal minimal theories at level $k=2$
equipped with the affine D--invariant.

It is of interest to consider the cohomology groups of the 3--fold
itself.  With the third Chern class $c_3=2h^3$ the Euler number of the
singular space is
\beq
\chi_s = \int c_3 =1
\eeq
and hence the Euler number of the resolved manifold is
\beq
\tchi=1-(2/2)+2\cdot  2 =4.
\eeq
Since the singular set is a sphere its resolution contributes just one
(1,1)--form and hence the second Betti number becomes $b_2=2$.
With  $\tchi=2(1+h^{(1,1)})-2h^{(2,1)}$ it follows that
$h^{(2,1)}=1$ and therefore the Hodge diamond becomes

\begin{center}
\begin{scriptsize}
\begin{tabular}{c c c c c c c}
  &  &          &1         &              &   &   \\
  &  &0         &          &0             &   &   \\
  &0 &          &2         &              &0  &   \\
0 &  &1         &          &1             &   &0  \\
  &0 &          &2         &              &0  &   \\
  &  &0         &          &0             &   &   \\
  &  &          &1         &              &   &   \\
\end{tabular}.
\end{scriptsize}
\end{center}

The case $n=3$  is exceptional because it leads to weights that are all
equal and hence involves a higher
dimensional manifold $\IP_5[3]$ that is smooth and hence the naive
expectation is that the construction breaks down.
This, however, is not the case, the observation being that the
polynomial to be considered is not of Fermat type and indeed has
a $\ZZ_2$ symmetry. Instead of considering the smooth manifold
$\IP_5[3]$ the space to use instead then is the orbifold
$\IP_5[3]/\ZZ_2$ and one needs to compare the
cohomology  groups of this space with the spectrum of K3
\beq
K3=\matrix{\IP_1\cr \IP_3\cr} \left[\matrix{2&0\cr 1&3\cr}\right],
\eeq
which consists of $h^{(0,0)}=h^{(2,2)}=h^{(2,0)}=h^{(0,2)}=1$ and
$h^{(1,1)}=20$, all other Hodge numbers are zero. Hence the Euler number
becomes $\chi(K3)=24$. The Euler  number for the noncritical manifold
\beq
\IP_5[3]
\eeq
is easily computed to be $\chi=27$. Since the manifold is K\"ahler,
$h^{(p,q)}=h^{(q,p)}$ and because of Poincar\'e duality, $b_p = b_{8-p}$.
Because the manifold has positive first Chern class, it follows from
Kodaira's vanishing theorem that $h^{(p,0)}=0$ for $p\neq 0$ and via
Lefshetz's hyperplane theorem it is known that below the middle dimension
all the cohomology is inherited from the ambient space and therefore
the  only  nonvanishing cohomology groups lead to $h^{(0,0)}=h^{(1,1)}=1$.
It can be shown that $h^{(3,1)}=h^{(1,2)}=1$ and therefore the only
remaining cohomology is in $H^{(2,2)}$. Since
\beq
\chi = 2(b_0+b_2) + 2+h^{(2,2)} = 6+h^{(2,2)},
\eeq
it follows that $h^{(2,2)}=21=20+1$. The $\ZZ_2$ orbifold of this
space introduces one further $(1,1)$--form and hence  we obtain
the spectrum of K3 plus one additional mode.
Thus the Hodge diamond splits as

{\scriptsize
\begin{center}
\begin{tabular}{c c c c c c c c c}
  &  &  &   &1           &            &   &   & \\
  &  &  &0  &            &0           &   &   & \\
  &  &0 &   &2           &            &0  &   & \\
  &0 &  &0  &            &0           &   &0  & \\
0 &  &1 &   &21          &            &1  &   &0 \\
  &0 &  &0  &            &0           &   &0  & \\
  &  &0 &   &2           &            &0  &   & \\
  &  &  &0  &            &0           &   &   &   \\
  &  &  &   &1           &            &   &   & \\
\end{tabular}
=
\begin{tabular}{c c c c c}
  &  &1  &  & \nn \\
  &0 &   &0 & \nn \\
1 &  &20  &  &1 \nn \\
  &0 &   &0 & \nn \\
  &  &1  &  & \nn \\
\end{tabular}
+
\begin{tabular}{c c c c c c c c c}
  &  &  &   &1           &            &   &   & \\
  &  &  &0  &            &0           &   &   & \\
  &  &0 &   &1           &            &0  &   & \\
  &0 &  &0  &            &0           &   &0  & \\
0 &  &0 &   &1           &            &0  &   &0 \\
  &0 &  &0  &            &0           &   &0  & \\
  &  &0 &   &1           &            &0  &   & \\
  &  &  &0  &            &0           &   &   &   \\
  &  &  &   &1           &            &   &   & \\
\end{tabular}.
\end{center}
}
This example is also useful because it indicates a generalization of the
considerations of the previous Section. The  surprising new feature of this
manifold is that even though the higher dimensional manifold did not have
any orbifold singularities it was nevertheless possible to decompose it in
such a way as to construct a Calabi--Yau manifold from it. This was possible
because the defining equation was not of Fermat type but involved coupling
between the fields. Because of this the manifold featured a new $\ZZ_2$
symmetry not present in the Fermat hypersurface and it is this new
symmetry that dictated how to perform the decomposition. This indicates that
even for smooth higher dimensional manifolds it is possible to relate them
to Calabi--Yau manifolds  once one moves away from the symmetric point.

Similar situations appear whenever
\beq
n=m+2
\eeq
in which case the defining equations in the ordinary projective space are
not of Fermat type and feature an additional $\ZZ_{m+1}$ symmetry that is
not induced by the ambient space.

The series (\ref{infser}) can be generalized to weighted critical manifolds
as is illustrated by the following polynomial
\beq
p=\sum_{i=1}^3 \left(x_i^3+x_iy_i^3\right) +
  \sum_{j=4}^5 x_i^6
\eeq
which originates from  the tensor model
$(16^3 \cdot 4)_{E^3 \otimes A^2}$ with central charge $c=12$ and
whose zero locus belongs to the configuration
\beq
\IP_{(6,4,6,4,6,4,3,3)}[18].
\eeq
The critical manifold derived from this 6--fold belongs to
the 4--dimensional configuration class
\beq
\matrix{\IP_2 \hfill\cr \IP_{(2,2,2,1,1)}\cr}
\left[\matrix{3&0\cr 2&6\cr}\right]
\eeq
which is indeed a Calabi--Yau deformation class.

Another infinite class of Calabi--Yau manifolds which are contained in
higher dimensional spaces of the type (\ref{spefam}) is furnished by the
series of manifolds
\beq
\IP_n[n+1],~~~~ n \in \IN
\eeq
which is related to the set of $(n+1)$--folds
\bea
\IP_{(1,1,....,1,\frac{n+1}{2},\frac{n+1}{2})}[n+1],
&~~~~~~(n+1)& even, \nn \\
\IP_{(2,2,....,2,n+1,n+1)}[2(n+1)], &~~~~~~(n+1)& odd.
\eea
For $(n+1)$ even the $\ZZ_{\frac{n+1}{2}}$--singularity set consists
of two points and therefore one obtains two copies of the Calabi--Yau
$\IP_n[n+1]$ and for odd $(n+1)$ the $\ZZ_{n+1}$ singular set is again
two points and the $\ZZ_2$ singular set is the Calabi--Yau.

The simplest case is $n=2$ for which the resolution of the
orbifold singularities of the noncritical 3--fold
\beq
\IP_{(2,2,2,3,3)}[6]
\lleq{counter}
leads to two independent Hodge numbers $h^{(1,1)}=4$, $h^{(2,1)}=2$
and hence the Hodge diamond
      contains {\it twice} the Hodge diamond of the torus, as it
must, according to the geometrical picture described above.
Similarly $\IP_{(2,2,2,2,2,5,5)}[10]$ leads
to two copies of the critical quintic.

Our construction is not restricted to the infinite series defined
in (\ref{infser}) or its weighted generalization
 as the next example illustrates.
A five--dimensional critical vacuum of higher codimension
is obtained by considering the Landau--Ginzburg potential
\beq
W= \sum_{j=1}^2 \left(u_i^3 + u_iv_i^2\right) +
    \sum_{i=3}^5 \left(u_i^3 + u_iw_i^3\right)
\eeq
which corresponds to the exactly solvable model
$(16^3\cdot 4^2)_{E_7^3 \otimes D^2}$ and is of a type different from the
ones discussed above. The nine--dimensional noncritical
manifold
\beq
\IP_{(3,2,3,2,3,2,3,3,3,3)}[9]
\eeq
leads, via its singularity structure, to the five--dimensional critical
manifold
\beq
\matrix{\IP_1\hfill  \cr \IP_2 \cr \IP_4\cr}
\left[\matrix{2 &0 &0\cr  0 & 3 &0\cr 1 & 1 & 3\cr}\right] .
\eeq
It is crucial that a polynomial was chosen which is not of
Brieskorn--Pham type for the
last four coordinates in the noncritical manifold.

There are of course other types of weighted $n$--folds which solve the
conditions (\ref{spefam}). An example that does not belong the class
above is
\beq
\IP_{(2,1,2,1,1,1)}[4] \ni
\left\{\sum_{i=1}^2 \left(x_i^2+x_iy_i^2\right)
                                    + x_5^4+x_6^4=0\right\}
\eeq
which leads to the critical surface
\beq
\matrix{\IP_1 \hfill \cr \IP_{(2,2,1,1)}\cr}
\left[\matrix{2&0 \cr 2&4\cr}\right].
\eeq
The cohomology of the noncritical 4--fold is easy to determine: since the
$\ZZ_2$--singular set consists of two points, the resolution of which
introduces two (1,1)--forms. Thus $h^{(1,1)}=3$.
With $c_4=25h^4$ one finds $\chi_s=25$ and hence
\beq
\chi=2(1+h^{(1,1)})-2=25-1+4= 28.
\eeq
Thus there appears, in general, not just one additional K\"ahler mode in
the spectrum of the higher dimensional manifold but more new fields whose
(possible) interpretation remains obscure at this point.

A final higher dimensional example involving a 4--dimensional critical
manifold of a different type starts from the  Landau--Ginzburg potential
\beq
W=\sum_{i=1}^3 \left(x_i^3+x_iy_i^3\right) +
  \sum_{j=4}^5 x_i^6
\eeq
with central charge $c=12$ which corresponds to the tensor model
$16_E^3 4_A^2$ and belongs to the configuration
\beq
\IP_{(6,4,6,4,6,4,3,3)}[18].
\eeq
The 4--dimensional manifold derived from this Landau--Ginzburg configuration
belongs to the configuration class
\beq
\matrix{\IP_2\hfill\cr \IP_{(2,2,2,1,1)}\cr}
\left[\matrix{3&0\cr 2&6\cr}\right]
\eeq
which is indeed a Calabi--Yau deformation class.

This brief discussion shows that many different types of special
Fano manifolds feature a structure which makes them amenable to our
construction, including a number of infinite series.

\vskip .2truein
\noindent
\section{Cohomology of Special Fano Spaces and Critical Spectra}

\noindent
In this Section I will describe in a more complete and coherent way the
cohomology of the higher dimensional manifolds and its relation to the
critical
spectrum.  The results described in the previous Sections suggest that the
spectrum of the critical vacuum ({\it not} described by a
Calabi--Yau manifold in general) is embedded in  the
cohomology of the higher dimensional space as:
\beq
H^{(p,q)}(V_{D_{crit}}) \subset H^{(p+Q-1,q+Q-1)}(\cM_{D_{crit}+2(Q-1)})
\eeq
or, put differently, there exists a projection $\pi$ such that
\beq
\pi \left(H^{(p,q)}(\cM_{D_{crit}+2(Q-1)})\right)
 = H^{(p-(Q-1),q-(Q-1))} (V_{D_{crit}}).
\eeq
What has been shown in \cite{rs93} and elaborated upon in the previous
Sections of the present paper
is that there exists a geometrical framework for this projection.

\noindent
There are two main results that follow from our geometrical construction.
First:
\begin{itemize}
\item
{\it The critical spectrum of a $c=D_{crit}$ vacuum of the N=1 spacetime
     supersymmetric string is determined by the cohomology carried by
     the singular sets of special Fano varieties of complex dimension
     $D_{crit}+2(Q-1)$ together with the monomials of the integrally
     charged subring of the polynomial (chiral) ring.}
\end{itemize}
 This is in contrast to Calabi--Yau manifolds where the spectrum is
described by the {\it resolution} of the singularities and the polynomial
ring.  From this an immediate consequence is that
\begin{itemize}
\item The cohomology of the higher dimensional ambient weighted projective
space does not contribute to the critical spectrum.
\end{itemize}

By using a combination of different techniques it is possible to compute
the Euler number and, more detailed, the cohomology groups of the
higher dimensional spaces.
First, one may recall that one way to compute the Euler number
of weighted complete intersection Calabi--Yau manifolds is to consider
them as coverings of the weighted spaces themselves.

Consider a Fermat configuration
\beq
\IP_{(k_1,\dots,k_{N+2})}[d] \ni
\left\{p=\sum_i z_i^{a_i}=0\right\}.
\eeq
Write this as
\beq
z_1^{a_1}=-z_2^{a_2}-\cdots - z_{N+2}^{a_{N+2}}.
\eeq
Clearly this describes an $a_1$--fold covering of
\beq
\IP_{(k_2,\dots,k_{N+2})}
\eeq
as long as $p_1$ does not vanish identically, the solution simply being
\beq
z_1 = \left(p_1(z_2,\dots,z_{N+2})\right)^{1/a_1},~~~~(z_2,\dots , z_{N+2})
\in
\IP_{(k_2,\dots,k_{N+2})}.
\eeq
When $p_1=0$, the only solution is $z_1=0$. Thus the space of solutions
of the equation $p=0$ becomes a $a_1$--fold covering of
$\IP_{(k_2,\dots,k_{N+2})}$ branched over the hypersurface
\beq
\IP_{(k_2,\dots,k_{N+2})}[d].
\eeq

This is an example of a general situation: if $M$ is an $n$--fold
covering of the space $N$ branched over a submanifold $B\in N$ then
the Euler number of the spaces involved are related via

\noindent
Riemann--Hurwitz:
$\chi(M) = n \chi(N) - (n-1)\chi(B)$.

This is the result when $M$ is a smooth manifold. In the
cases at hand however the manifolds are singular in general and
Riemann--Hurwitz computes the Euler number for the singular manifold
which has to be
resolved: if the manifold has orbifold singularities $S_i$ with respect
to cyclic groups $\ZZ_{n_i}$ of order $n_i$ then the Euler number of the
resolved manifold $\tM$ is given by
\bea
\chi(\tM) &=& \chi(M) + \sum_i (n_i-1) \chi(S_i) \nn \\
          &=& n\chi(N) - (n-1)\chi(B) + \sum_i (n_i-1) \chi(S_i).
\eea

A simple but important example, as will become presently, is the
3--fold
\beq
\IP_{(2,2,2,3,3)}[6] \ni
\left\{ \sum_i x_i^3 +\sum y_j^2 =0\right\}.
\eeq
The singular sets consist of the singular curve
\beq
\ZZ_2: ~~~\IP_2[3],
\eeq
which describes the torus, and two $\ZZ_3$--singular points
\beq
\ZZ_3: ~~~\IP_1[2]=2~{\rm pts}.
\eeq
The novel feature of this manifold is the fact that the fiber, which plays
such an important r\^{o}le in the construction of Section 3,  collapses to
two points and hence the critical manifold becomes
\beq
\matrix{\IP_2 \cr \IP_1\cr}
\left[\matrix{3&0\cr 0&2\cr}\right],
\eeq
i.e. it consists of two copies of the torus.

The noncritical spectrum can be obtained as follows: first one computes
via Riemann--Hurwitz the Euler number, which turns out to be
\beq
\chi_s = 2\cdot 4 - \chi(\IP_{2,2,2,3)}[6])
       = 8 - 2\cdot 3 + \chi(\IP_2[3]) = 2.
\eeq
Resolving the singular space leads to
\beq
\chi = -\chi_s +\sum_i (n_i-1)\chi(S_i)
     =  2 + 2\cdot 2 = 6.
\eeq
This result can be checked with the standard resolution
formula for the Euler number, using  $c_3=8h^3$,
\beq
\chi = \frac{8\cdot 6}{8\cdot 9} - \frac{2}{3} + 3\cdot 2 = 6.
\eeq

Resolving these singularities leads to the surface with the local fibering
\beq
\IP_2[3] \times \IP_1,~
\eeq
which, via K\"unneth, leads to one additional (1,1)--form wheras
the resolution
of the two $\ZZ_3$ points introduces 2 additional (1,1)--forms.
Using the result for the Euler number then leads to
\beq
\chi=6=2(1+4)-2h^{(2,1)}
\eeq
and hence to $h^{(4,1)}=4$.

Thus the Hodge diamond of the 3--fold is completely determined and
leads to  (including the 1 monomial of charge 1)

{\scriptsize
\bea
\begin{tabular}{c c c c c c c c c c c c c c c}
  &  &  &   &   &  &    &1  &    &   &   &   &   &   &   \nn \\
  &  &  &   &   &  &0   &   &0   &   &   &   &   &   &   \nn \\
  &  &  &   &   &0 &    &4  &    &0  &   &   &   &   &   \nn \\
  &  &  &   &0  &  &2   &   &2   &   &0  &   &   &   &   \nn \\
  &  &  &   &   &0 &    &4 &    &0  &   &   &   &   &   \nn \\
  &  &  &   &   &  &0   &   &0   &   &   &   &   &   &   \nn \\
  &  &  &   &   &  &    &1  &    &   &   &   &   &   &   \nn \\
\end{tabular}.
\eea
}

This simple example is interesting  because it features a new phenomenon which
also occurs for critical vacua in the physical dimension, i.e. for special
Fano folds that lead to Calabi--Yau manifolds of dimension $dim_{\IC} M=3$.
The important consequence is that the higher dimensional analog of the
critical holomorphic $(D_{crit},0)$ is not always 1--dimensional, a fact for
which our construction provides a simple geometrical explanation, the reason
being that multiple copies of the critical manifold can be embedded in the
higher dimensional space.

An example that leads to a phenomenological interesting
string vacuum is furnished by
\beq
\IP_{(2,2,2,3,3,3,3)}[9] \ni
\left\{ \sum_i \left(x_i^3y_i +y_i^3\right) +x_4^3 =0\right\}.
\eeq
There are two singular surfaces embedded in this manifold, the projective
plane and the cubic surface
\beq
\ZZ_2: ~~~\IP_2 :~~~~
{\scriptsize
\begin{tabular}{c c c c c}
  &  &1  &  & \nn \\
  &0 &   &0 & \nn \\
0 &  &1  &  &0 \nn \\
  &0 &   &0 & \nn \\
  &  &1  &  & \nn \\
\end{tabular}
}, ~~~~~~
\ZZ_3: ~~~\IP_3[3] :~~~~
{\scriptsize
\begin{tabular}{c c c c c}
  &  &1  &  & \nn \\
  &0 &   &0 & \nn \\
0 &  &7  &  &0 \nn \\
  &0 &   &0 & \nn \\
  &  &1  &  & \nn \\
\end{tabular}
}.
\eeq
When resolved, these lead to the 3--folds
\beq
\IP_2 \times \IP_1,~~~~\IP_3[3]\times \IP_1
\eeq
respectively.
Using K\"unneth's formula these 3--folds lead to the following Hodge diamond
of the resolved higher dimensional Fano manifold of special type
{\scriptsize
\begin{center}
\begin{tabular}{c c c c c c c c c c c c c c c}
  &  &  &   &   &  &    &1  &    &   &   &   &   &   &   \nn \\
  &  &  &   &   &  &0   &   &0   &   &   &   &   &   &   \nn \\
  &  &  &   &   &0 &    &3  &    &0  &   &   &   &   &   \nn \\
  &  &  &   &0  &  &0   &   &0   &   &0  &   &   &   &   \nn \\
  &  &  &0  &   &0 &    &9  &    &0  &   &0  &   &   &   \nn \\
  &  &0 &   &1  &  &35  &   &35   &   &1  &   &0  &   &   \nn \\
  &  &  &0  &   &0 &    &9  &    &0  &   &0  &   &   &   \nn \\
  &  &  &   &0  &  &0   &   &0   &   &0  &   &   &   &   \nn \\
  &  &  &   &   &0 &    &3  &    &0  &   &   &   &   &   \nn \\
  &  &  &   &   &  &0   &   &0   &   &   &   &   &   &   \nn \\
  &  &  &   &   &  &    &1  &    &   &   &   &   &   &   \nn \\
\end{tabular}.
\end{center}
}
\noindent
Thus the Euler number becomes
\beq
\chi=2(b_0-b_1+b_2-b_3+b_4)-b_5 = -46.
\eeq

The critical manifold
\beq
\matrix{ \IP_2 \cr \IP_3\cr} \left[\matrix{3&0\cr 1&3\cr}\right]
\lleq{3gcicy}
which is embedded in the special Fano 5--fold is of some interest because it
is the starting point of the construction of a 3--generation Calabi--Yau
manifold introduced  in \cite{rs87}.

This example is misleading in one respect however: all of the
critical generations are parametrized by the chiral ring itself,
i.e. the monomials of charge 1. This translates into the fact that
all complex deformations of the critical Calabi--Yau space (\ref{3gcicy})
which derives from this Fano fold via our construction are indeed
parametrized by the monomials of the polynomials.
In many cases however these monomials fail to parametrize all complex
deformations, a simple example being furnished by the complete intersection
Calabi--Yau manifold \cite{gh87}
\beq
\matrix{ \IP_1 \cr \IP_4\cr} \left[\matrix{2&0\cr 1&4\cr}\right].
\eeq
This critical manifold can be derived from the higher dimensional space
\beq
\IP_{(2,2,2,2,2,3,3)}[8] \ni
   \left\{ \sum_{i=1}^2 \left(x_i^4 + x_iy_i^2\right)
                              + x_3^4 + x_4^4 + x_5^4 = 0 \right\},
\eeq
the analysis of which shows that indeed the `missing' complex deformations,
not parametrized by the complex monomials, originate from the singularities
of the higher dimensional space.

It should be realized that the computation of the critical spectrum from the
higher dimensional space does {\it not} depend on the ability to find an
embedding of a 3--dimensional Calabi--Yau space. Consider, e.g.
\beq
\IP_{(1,1,2,2,2,2,2)}[6] \ni
\{z_1^6 +z_2^6 + \sum_i z_i^3=0\}
\eeq
for which the Riemann--Hurwitz formula leads to the singular Euler number
\beq
\chi_s = 3\cdot 6 - 2 \chi(\IP_{1,1,2,2,2,2)}[6])
       = 18 -2\{ 3\cdot 5 - 2\chi(\IP_{(1,1,2,2,2)}[6])\}
       = \cdots = -132
\eeq
which in turn leads to the resolved Euler number
\beq
\chi =-132 + \chi(\IP_4[3]) =-138
\eeq
after blowing up the singular cubic 3--fold.

{}From the singular cubic 3--fold, which has Euler number $-6$ one obtains
immediately that the number of critical (1,1)--forms is 1 and that
there are 5 additional critical (2,1)--modes coming from the
singularity, in addition to the 68 monomial deformations.
Thus for the total critical spectrum one obtains
\beq
(h^{(1,1)}_{crit}, h^{(2,1)}_{crit})(\IP_{(1,1,2,2,2,2,2)}[6]) = (1,73)
\eeq
which is the correct string spectrum.

It is also of interest to compute the cohomology of the noncritical manifold.
In order to obtain the cohomology groups of the ambient space it is only
necessary to compute, via K\"unneth the Hodge diamond of the bundle
$\IP_3[4]\times \IP_1$, leading to one additional (1,1)--form. Using the
result for the Euler number one finds
\beq
-138=\chi=2(b_0+b_2+b_4)-b_5= 10- 2h^{(4,1)}-146
\eeq
and hence $h^{(4,1)}=1$, as expected. Thus the Hodge diamond
{\scriptsize
\bea
\begin{tabular}{c c c c c c c c c c c c c c c}
  &  &  &   &   &  &    &1  &    &   &   &   &   &   &   \nn \\
  &  &  &   &   &  &0   &   &0   &   &   &   &   &   &   \nn \\
  &  &  &   &   &0 &    &2  &    &0  &   &   &   &   &   \nn \\
  &  &  &   &0  &  &0   &   &0   &   &0  &   &   &   &   \nn \\
  &  &  &0  &   &0 &    &2  &    &0  &   &0  &   &   &   \nn \\
  &  &0 &   &1  &  &73 &   &73 &   &1  &   &0  &   &   \nn \\
  &  &  &0  &   &0 &    &2  &    &0  &   &0  &   &   &   \nn \\
  &  &  &   &0  &  &0   &   &0   &   &0  &   &   &   &   \nn \\
  &  &  &   &   &0 &    &2 &    &0  &   &   &   &   &   \nn \\
  &  &  &   &   &  &0   &   &0   &   &   &   &   &   &   \nn \\
  &  &  &   &   &  &    &1  &    &   &   &   &   &   &   \nn \\
\end{tabular}
\eea
}
is obtained.

The general Hodge--diamond for odd--dimensional manifolds,
$dim_{\IC} M = (2p+1)$, is
{\small
\begin{center}
\begin{tabular}{c c c c c c c c c c c c c c c c c}
   &  &  &        &  &  &  &  &1           &  &   &   &   &       &  & & \\
   &  &  &        &  &  &  &0 &            &0 &   &   &   &       &  & & \\
   &  &  &        &  &  &0 &  &{\scriptsize $h^{(1,1)}$}
                                           &  &0  &   &   &       &  & & \\
   &  &  &        &  &0 &  &0 &            &0 &   &0  &   &       &  & & \\
   &  &  &        &0 &  &0 &  &{\scriptsize $h^{(2,2)}$}
                                           &  &0  &   &0  &       &  & & \\
   &  &  &0       &  &0 &  &0 &            &0 &   &0  &   &0      &  & & \\
   &  &  &$     $ &  &  &  &  &$\vdots$    &  &   &   &   &       &  & & \\
   &  &  &$     $ &  &  &  &  &$\vdots$    &  &   &   &   &       &  & & \\
   &0 &  &$\cdots$&  &  &0 &  &{\scriptsize $h^{(p,p)}$}
                                           &  &0  &   &   &$\cdots$
                                                                  &  &0& \\
  0&$\cdots$
       &
         &0       &  &$n$ &  &{\scriptsize $h^{(p+1,p)}$}
                              &            &{\scriptsize $h^{(p,p+1)}$}
                                              &   &$n$  &
                                                          &0      &
                                                                &$\cdots$
                                                                   &0  \\
\end{tabular},
\end{center}
}

\noindent
whereas for even--dimensional manifolds, $\dim_{\IC}=2p$, it is of the form
{\small
\begin{center}
\begin{tabular}{c c c c c c c c c c c c c c c c c}
   &  &  &        &  &  &  &  &1           &  &   &   &   &       &  & & \\
   &  &  &        &  &  &  &0 &            &0 &   &   &   &       &  & & \\
   &  &  &        &  &  &0 &  &{\footnotesize $h^{(1,1)}$}
                                           &  &0  &   &   &       &  & & \\
   &  &  &        &  &0 &  &0 &            &0 &   &0  &   &       &  & & \\
   &  &  &        &0 &  &0 &  &{\footnotesize $h^{(2,2)}$}
                                           &  &0  &   &0  &       &  & & \\
   &  &  &0       &  &0 &  &0 &            &0 &   &0  &   &0      &  & & \\
   &  &  &$     $ &  &  &  &  &$\vdots$    &  &   &   &   &       &  & & \\
   &  &  &$     $ &  &  &  &  &$\vdots$    &  &   &   &   &       &  & & \\
  0&
      &$\cdots$
         &        &0 &
                        &$n$
                           &  &{\footnotesize $h^{(p,p)}$}
                                           &  &$n$&
                                                      &0   &     &$\cdots$
                                                              &        &0 \\
\end{tabular}.
\end{center}
}

Even though, from an abstract cohomological point of view, it is not known how
to perform the projection of the special Fano Hodge diamond to the critical
stringy part, our construction provides a geometrical prescription which
does provide
a universal tool to perform the decomposition of the ambient cohomology into a
stringy and a remaining part. One universal feature which follows from this
geometrical construction is the fact
that in the transition from the higher dimensional spectrum to the
critical spectrum, described by the generations and antigenerations,
the cohomology of the ambient projective space does not contribute.

\vskip .2truein
\noindent
\section{When $h^{D_{crit}+Q-1,Q-1)} >1$: a Degeneration Phenomenon.}

\noindent
It is tempting to try a cohomological definition of the special class of
Fano folds (\ref{spefam}) as a class of spaces for which there exists
precisely one analog of the holomorphic $(D_{crit},0)$--form which exists in
critical string vacua. This however would be fallacious.

 Consider the real life example
\beq
\IP_{(2,2,2,2,2,5,5)}[6] \ni
\left\{ \sum_i x_i^3 +\sum y_j^2 =0\right\}
\eeq
whose singular sets consist of the 3--fold
\beq
\ZZ_2: ~~~\IP_4[5],~~~~\chi=-200
\eeq
and the two points
\beq
\ZZ_5: ~~~\IP_1[2]=2~{\rm pts}.
\eeq
Going back to the discussion of Sections 3 and 4, we see that the fiber
collapses down to two points in this example and
the critical manifold becomes
\beq
\matrix{\IP_4 \cr \IP_1\cr}
\left[\matrix{5&0\cr 0&2\cr}\right],
\eeq
i.e. it consists of two copies of the quintic.

This {\it degeneration phenomenon} also shows up through an analysis of the
Hodge structure of the manifold.
The noncritical spectrum can be obtained as follows: first one computes
via Riemann--Hurwitz the Euler number, which turns out to be
\beq
\chi_s = 2\cdot 6 - \chi(\IP_{2,2,2,2,2,3)}[10]) =-198
\eeq
which, when resolved, leads to
\beq
\chi = -\chi_s +\sum_i (n_i-1)\chi(S_i)
     = -198 + (-200) + 4\cdot 2 = -390.
\eeq
This result can be checked with the  resolution
formula for the Euler number: with $c_5=-7968h^5$ it follows that
\beq
\chi=- \frac{498}{5} - \frac{-200}{2}+2(-200) -\frac{2}{5} + 5\cdot 2 =-390.
\eeq

In more detail, by resolving these singularities, one is locally lead to
the 4--fold
\beq
\IP_4[5] \times \IP_1
\eeq
which, via K\"unneth, leads to one additional (1,1)--form
whereas the resolution of the two $\ZZ_5$ points introduces 2
additional (1,1)--forms.  Using the above result for the Euler number
then leads to
\beq
\chi=-390=2(1+4+4)-2h^{(4,1)}-404
\eeq
and hence to $h^{(4,1)}=2$.

Thus the Hodge diamond of the 5--folds is completely determined and
leads to
{\scriptsize
\bea
\begin{tabular}{c c c c c c c c c c c c c c c}
  &  &  &   &   &  &    &1  &    &   &   &   &   &   &   \nn \\
  &  &  &   &   &  &0   &   &0   &   &   &   &   &   &   \nn \\
  &  &  &   &   &0 &    &4  &    &0  &   &   &   &   &   \nn \\
  &  &  &   &0  &  &0   &   &0   &   &0  &   &   &   &   \nn \\
  &  &  &0  &   &0 &    &4  &    &0  &   &0  &   &   &   \nn \\
  &  &0 &   &2  &  &202 &   &202 &   &2  &   &0  &   &   \nn \\
  &  &  &0  &   &0 &    &4  &    &0  &   &0  &   &   &   \nn \\
  &  &  &   &0  &  &0   &   &0   &   &0  &   &   &   &   \nn \\
  &  &  &   &   &0 &    &4 &    &0  &   &   &   &   &   \nn \\
  &  &  &   &   &  &0   &   &0   &   &   &   &   &   &   \nn \\
  &  &  &   &   &  &    &1  &    &   &   &   &   &   &   \nn \\
\end{tabular}
\eea
}
(including the 101 monomials of charge unity).

This example is important because it shows two things: the first
is that also for the physically relevant situation of Calabi--Yau manifolds
of three complex dimensions it can happen that  $h^{(D_{crit}+Q-1,Q-1)}>1$!
This shows that, contrary to what one might expect, the higher dimensional
analog of the critical holomorphic $(D_{crit},0)$ is not always 1--dimensional,
the simple reason being that multiple copies of the critical
manifold can be embedded in the higher dimensional space.
The second point is that, contrary to the case of smooth
noncritical manifolds such as $\IP_8[3]$ and $\IP_{(1,1,1,1,1,1,2)}[4]$,
singular spaces have a much more complicated cohomology.
Hence such manifolds lead, in general,
to many more states than those comprising the critical spectrum.

\vskip .2truein
\noindent
\section{Splitting and Contraction: Many Fano Fold Embedding of Critical Vacua}

\noindent
In this Section it is pointed out that the relation between higher dimensional
manifolds and critical vacua is not 1--1. Instead many higher dimensional
manifolds are related to one and the same critical vacuum. In general even
an infinite number.

One way to see this, is via the process of splitting and contraction
introduced in \cite{cdls88}.  The discussion of ref. \cite{cdls88} is
restricted to Calabi--Yau manifolds embedded in products of ordinary
projective spaces but it readily generalizes to the more general framework
of weighted projective spaces.

It is useful to start with an example. Consider the 7--fold
\beq
\IP_{(1,1,10,10,2,2,2,2,6)}[12] \ni
\left\{ \sum_{i=1}^2 \left( x_i^2y_i + y_iz_i +z_i^6\right)
               + z_3^6+z_4^6+z_5^2 =0 \right\}
\eeq
where $(x_1,x_2,y_1,y_2,z_1,z_2,...,z_5)$ are the coordinates of the
ambient weighted space. This space leads, via an iterative construction,
as described in the previous Section, to the critical manifold
\beq
\matrix{\IP_{(1,1)}\hfill\cr \IP_{(1,1)}\hfill\cr \IP_{(1,1,1,1,3)}\cr}
\left[\matrix{2&0&0\cr 1&1&0\cr 0&1&6\cr}\right]
\lleq{split}
with the by now familiar split of the polynomial.

Using the same analysis as described in \cite{cdls88} it is however
easy to see that this critical manifold is in fact isomorphic to the
Calabi--Yau manifold
\beq
\matrix{\IP_{(1,1)} \hfill \cr \IP_{(1,1,1,1,3)}\cr}
\left[\matrix{2&0\cr 1&6\cr}\right]
\lleq{example}
described by the polynomials
\bea
p_1&=&x_1^2y_1+x_2^2y_2~, \nn \\
p_2&=&y_1^6+y_2^6+y_3^6+y_4^6+y_5^2
\eea
which can be derived via our construction from the higher dimensional
manifold
\beq
\IP_{(5,5,2,2,2,2,6)}[12]\ni
       \left\{\sum_{i=1}^2\left( x_i^2y_i + y_i^6\right)
                         +y_3^6 + y_4^6 +y_5^2=0 \right\}
\eeq
where $(x_1,x_2,y_1,y_2,...,y_5)$ are the homogeneous coordinates of the
6--dimensional weighted space. Manifold (\ref{split}), in fact, is the
so--called split \cite{cdls88} of the manifold
(\ref{example}) and in special circumstances, which apply in the present case,
the splitting or contraction process does not change
the theory but rather provides different representations of the same
manifold. In such a situation one is dealing with a split that is said to
be {\it ineffective}.

It turns out  that the manifold (\ref{example}) can be represented
via ineffective
splits in an infinite number of ways as
\beq
\matrix{\IP_{(1,1)} \hfill \cr \IP_{(1,1,1,1,3)}\cr}
\left[\matrix{2&0\cr 1&6\cr}\right]
\longrightarrow
\matrix{\IP_{(1,1)}\hfill\cr \IP_{(1,1)}\hfill\cr \IP_{(1,1,1,1,3)}\cr}
\left[\matrix{2&0&0\cr 1&1&0\cr 0&1&6\cr}\right]
\longrightarrow
\matrix{\IP_{(1,1)}\hfill\cr \IP_{(1,1)}\hfill\cr\IP_{(1,1)}\hfill \cr
 \IP_{(1,1,1,1,3)}\cr}
\left[\matrix{2&0&0&0\cr 1&1&0&0\cr 0&1&1&0\cr 0&0&1&6}\right]
\longrightarrow
\cdots
\eeq
This chain of different representations of the same Calabi--Yau 3--fold
can be derived from a corresponding sequence of higher dimensional
special Fano spaces
\beq
\IP_{(5,5,2,2,2,2,6)}[12] \longrightarrow
\IP_{(1,1,10,10,2,2,2,2,6)}[12] \longrightarrow
\IP_{(5,5,2,2,10,10,2,2,2,2,6)}[12] \longrightarrow \cdots
\eeq
i.e. the infinite sequence of
\beq
\IP_{(5,5,2,2,10,10,2,2,10,10,...,2,2,10,10,2,2,6)}[12]
\eeq
where the part $(2,2,10,10)$ occurs $k$ times, and
\beq
\IP_{(1,1,10,10,2,2,10,10,2,2,...10,10,2,2,2,2,6)}[12],
\eeq
where $(10,10,2,2)$ occurs $k$ times. The burden of this observation is that
a given string vacuum can be derived from an infinite sequence of special
Fano manifolds with ever increasing dimension.

This example easily generalizes to a class of spaces \cite{rs89} which was
mentioned already in Section 3.  Consider manifolds embedded in
\beq
\matrix{\IP_{(1,1)} \hfill\cr \IP_{(k_1,k_1,k_3,k_4,k_5)}\cr}
\left[\matrix{2&0\cr k_1&k\cr}\right],
\eeq
where $k=k_1+k_3+k_4+k_5$. These spaces can be split
into the infinite sequences
\beq
\matrix{\IP_{(1,1)} \hfill \cr \IP_{(k_1,k_1,k_3,k_4,k_5)}\cr}
\left[\matrix{2&0\cr k_1&k\cr}\right]
\longrightarrow
\matrix{\IP_{(1,1)} \hfill \cr \IP_{(1,1)}\hfill \cr
 \IP_{(k_1,k_1,k_3,k_4,k_5)}\cr}
\left[\matrix{2&0&0\cr 1&1&0\cr 0&k_1&k\cr}\right]
\longrightarrow
\matrix{\IP_{(1,1)} \hfill \cr \IP_{(1,1)}\hfill\cr\IP_{(1,1)}\hfill \cr
 \IP_{(k_1,k_1,k_3,k_4,k_5)}\cr}
\left[\matrix{2&0&0&0\cr 1&1&0&0\cr 0&1&1&0\cr 0&0&k_1&k}\right]
\longrightarrow
\cdots
\eeq
with the corresponding sequence of higher dimensional spaces given by
\bea
\IP_{(k-k_1,k-k_1,2k_1,2k_1,2k_3,2k_4,2k_5)}[2k]
\longrightarrow
\IP_{(k_1,k_1,2(k-k_1),2(k-k_1),2k_1,2k_1,2k_3,2k_4,2k_5)}[2k]
\longrightarrow \cdots
\eea
i.e. they belong to the set of spaces of type
\beq
\IP_{((k-k_1),(k-k_1),2k_1,2k_1,2(k-k_1),2(k-k_1),....,
 2k_1,2k_1,2k_3,2k_4,2k_5)}[2k]
\eeq
where the part $(2k_1,2k_1,2(k-k_1),2(k-k_1))$ occurs $p$ times, and
\beq
\IP_{(k_1,k_1,2(k-k_1),2(k-k_1),
2k_1,2k_1,...,2k_3,2k_4,2k_5)}[2k],
\eeq
where $(2(k-k_1),2(k-k_1),2k_1,2k_1)$ occurs $p$ times.

\vskip .2truein
\noindent
\section{Topological Relations between Fano Varieties}

It is the purpose of this short Section to point out that certain topological
identities, such as those employed in \cite{cdls88}, make it possible to apply
our embedding construction of Section 3 to Calabi--Yau manifolds for which
naively an analysis along the present lines appears impossible.
The reason for this
is that these topological identities allow us to represent complete
intersection Calabi--Yau manifolds in many different ways.

Consider the codimension--four Calabi--Yau manifold
\beq
\matrix{\IP_1 \cr \IP_2\cr \IP_2 \cr \IP_2\cr}
\left[\matrix{2 &0 &0 &0\cr
              1 &2 &0 &0\cr
              0 &1 &2 &0\cr
              0 &0 &1 &2\cr}\right]
\lleq{boggle}
with the defining polynomials
\beq
\begin{array}{r l r l}
p_1 &=~ \sum_{i=1}^2 u_i^2v_i, & p_2 &=~ \sum_{i=1}^3 v_i^2w_i,  \\
p_3 &=~ \sum_{i=1}^3 w_i^2x_i, & p_4 &=~ \sum_{i=1}^3 x_i^2
\end{array}
\lleq{bogglepolly}
which can be derived from
\beq
\IP_{(5,5,6,6,6,4,4,4,8,8,8)}[16]
\eeq
as discussed in Section 4.

This manifold can be represented in rather different  ways by using a
well--known representation of the sphere as a quadric in the
projective plane
\beq
S^2 = \IP_1 = \IP_2[2].
\eeq
Using this,  the manifold (\ref{boggle}) can be rewritten as
\beq
\matrix{\IP_1 \cr \IP_2\cr \IP_2 \cr \IP_1\cr}
\left[\matrix{2 &0 &0 \cr
              1 &2 &0 \cr
              0 &1 &2 \cr
              0 &0 &2 \cr}\right].
\eeq
Furthermore one can use the surface identity \cite{cdls88}
\beq
\matrix{\IP_1 \cr \IP_2\cr} \left[\matrix{2 \cr 1\cr}\right]
=\matrix{\IP_1 \cr \IP_1}
\eeq
via the application rule
\beq
\matrix{\IP_1\cr \IP_2\cr X\cr }
\left[\matrix{2&0\cr 1&a\cr 0&M\cr}\right]
=\matrix{\IP_1 \cr \IP_1\cr X\cr}
\left[\matrix{a\cr a\cr M\cr}\right]
\eeq
to obtain the representation
\beq
\matrix{\IP_1 \cr \IP_1\cr \IP_1 \cr\IP_2\cr}
\left[\matrix{2 &0 \cr
              2 &0 \cr
              0 &2 \cr
              1 &2 \cr}\right]
\eeq
whose defining polynomials are some deformations of
\begin{eqnarray}
p_1&=&\sum_i u_i^2 v_i^2 y_i \nn \\
p_2&=&\sum_i x_i^2 y_i^2 .
\end{eqnarray}
The natural ansatz for the higher dimensional polynomial
\beq
p=p_1+p_2
\eeq
clearly does not have an isolated singularity and hence does not
have an interpretation as the defining equation of a quasihomogeneous
hypersurface in some weighted projective space.  In fact it is not even
possible to assign weights to the variables such that a weighted projective
space could be specified in which the purported hypersurface is supposed to
live. But because this manifold is diffeomorphic to the
variety (\ref{boggle}) it does admit, via this equivalence, an embedding
into a higher dimensional space.

\vskip .2truein
\noindent
\section{Landau--Ginzburg Theories, Special Fano Manifolds, and
        Critical Vacua}

\noindent
The basic idea in \cite{rs93} is that it is the singularity structure of the
higher dimensional special Fano manifolds (and the polynomial ring)
that determines
the string physics. This idea also leads to a new way of looking at the
Calabi--Yau/Landau--Ginzburg connection as will be discussed presently.

The starting point of the mean field description of N$=$2
superconformal theory is an N$=$2 supersymmetric field theory
in terms of a set of chiral superfields
$\Phi_i(z,\bz,\th^{\pm},\bth^{\pm})$, defined on the superspace
parametrized by coordinates $(z,\bz,\theta^{\pm},\bth^{\pm})$
via the constraints
\beq
D^+ \Phi_i=0=\bD^+  \Phi_i
\eeq
where
\beq
D^{\pm} = \frac{\del}{\del \th^{\pm}}+\th^{\mp} \del_z
\eeq
(and similarly for the right moving sector).

The action for these fields will be chosen to be of the form
\beq
\int d^2z d^2\th d^2\bth~ K(\Phi_i,\bPhi_i) +
\int d^2z d^2\th~ W(\Phi_i) + c.c.                      \nn
\eeq
where $K$ is the K\"ahler potential and $W$ is the superpotential
which is assumed to be quasihomogeneous. This means that there exists
a set of weights $k_i>0$ such that under
$\Phi_i \rightarrow \l^{k_i}\Phi_i$
the superpotential scales as
\beq
W(\l^{k_i} \Phi_i) = \l^d W(\Phi_i)
\eeq
for some integer $d$.

It was the insight of Martinec \cite{m89} and Vafa and
Warner \cite{vw89} that such Landau--Ginzburg theories are useful
for the understanding of string
vacua and that much information about such groundstates is already
encoded in the superpotential. The point that was emphasized by
these authors is that since the K\"ahler potential contains
only irrelevant and marginal operators the superpotential $W$
already determines essential properties of the theory.

The important structure determined by the superpotential
is the ring of monomials that one can build
with the order parameters $\Phi_i$. This ring is defined by all
monomials modulo those that are generated by the ideal $\cI$ defined
by the partial derivatives $\del_i W$ of the superpotential with
respect to the order fields $\Phi_i$
\beq
\cR = \frac{\IC[\Phi_i]}{\cI[\del_i W]}.
\eeq
Since this ring is finite dimensional, its dimension given by the
Milnor number
\beq
\mu = \prod_i \left(\frac{k_i}{d}-1\right),
\eeq
there must exist a  monomial of maximal degree. This monomial is important
because it corresponds to the field of maximal weight $\Delta = c/3$ that
always exists in an $N=2$ supersymmetric conformal field theory.  Using the
fact of Arnold that the maximal degree obtainable is determined by
\beq
d_{max} = \sum_{i=1}^N \left(1-2q_i\right),
\eeq
it follows \cite{m90} that the central charge of such a Landau--Ginzburg
theory is
\beq
c=3\sum_{i=1}^N \left(1-2q_i\right),
\eeq
where $q_i=k_i/d$.

The concept of Landau--Ginzburg theories of course allows to describe
compactifications of the heterotic string not only from ten dimensions
down to four dimensions by using a Landau--Ginzburg theory of central charge
nine, but also allows to describe compactifications down to six dimensons via
theories with central charge six, or to eight spacetime dimensions
with theories of central charge three. The critical dimension of the internal
theory, if it has a spacetime interpretation, in all these case is
\beq
D_{crit} = c/3.
\eeq
In the context of Landau--Ginzburg theories the integer $Q$ introduced in
Section 2 is simply the total charge and one can rewrite the central charge as
\beq
N=D_{crit}+2Q
\eeq
thus determining the number of scaling fields in terms of the
critical dimension and the total charge.

If $D_{crit}$ is an integer then clearly $2Q \in \IN$ and it is
always possible to add an additional trivial field in order to obtain
integer total charge $Q\in \IN$.
  As mentioned already  $Q$ has the geometrical interpretation
of a codimension: it describes into how many polynomial constraints the
superpotential must split in order to define a manifold of critical
dimension $D_{crit}$. Given the number of constraints, the critical
dimension, and the number of fields in the theory, it follows that
the number of projective scales is
\beq
\cF = N - D_{crit} - Q
\eeq
and therefore
\beq
\cF =Q.
\eeq

Consider now the path integral
\beq
Z=\int \prod_i D\Phi_i \prod_j D\Psi_j
e^{-\int d^2zd^4\theta K(\Phi_i, {\bar \Phi_i},\Psi_j, {\bar \Psi}_j)
               +\int d^2zd^2\theta W(\Phi_i,\Psi_i)+c.c.}
\lleq{pint}
with the superpotential
\beq
W(\Phi_i,\Psi_i) = \sum_{i=1}^3\Psi_i^3\Phi_i + \sum_{j=1}^4 \Phi_j^3.
\lleq{mypot}
A priori there is no reason to ask for a geometrical description of the
conformal
fixed points of this model. However this theory in itself does not
describe a string ground state, the missing ingredient being the GSO
projection.
Implementation of this projection in this context amounts to compactifying the
affine variety and hence the question arises what spacetime interpretation the
resulting theory affords.

This is were hindsight comes in. By computing the spectrum of the
GSO projected Landau--Ginzburg theory \cite{v89} and comparing the result
with known Calabi--Yau manifolds it becomes obvious that the theory
(\ref{mypot}) should related to a manifold embedded in the configuration
\cite{rs87}
\beq
 \matrix{\IP_2\cr \IP_3\cr}
 \left[\matrix{3&0\cr 1&3\cr}\right]
\lleq{smf}
and defined by the polynomial constraints
\bea
p_1 &=& \sum_{i=1}^3 y_i z_i^3,  \nn \\
p_2 &=& \sum_{j=1}^4 y_i^3
\eea
where the superpotential $W$ has been split into two parts.
This is supported by an analysis of the underlying exactly solvable model of
the potential (\ref{mypot}). Indeed, it was shown by Gepner
\cite{g87a}\cite{g87b},   even before the appearance of
\cite{m89} \cite{vw89}, through a detailed analysis of
the spectrum and the symmetries that the manifold (\ref{smf}) is related
to the GSO--projected tensor model
\beq
(1\cdot 16^3)_{A_2 \otimes E_7^3}
\eeq
of N$=$2 exactly solvable minimal models at level 1 and level 16, endowed
with the diagonal modular invariant and the $E_7$ invariant respectively.
The results of \cite{m89}\cite{vw89} support this connection.

Knowing what to shoot for then helps in massaging the path integral above into
a form which suggests a relation to (\ref{smf}). The idea \cite{gvw89} is to
obtain some $\delta$--function constraints, restricting the variables to lie
on a hypersurface. To achieve this, define new coordinates
\beq
\xi_1 = \Phi_1^3,~~\xi_i= \frac{\Phi_i}{\Phi_1},~i=2,...,4,
\xi_5 = \Phi_1 \Psi_1^3, ~~\xi_6= \frac{\Psi_2}{\Psi_1}, ~~
                           \xi_7= \frac{\Psi_3}{\Psi_1}.
\eeq
As emphasized in \cite{gvw89} it is important for this coordinate
change to lead
to a constant Jacobian. Because of this we can write (neglecting the kinetic
term)
\bea
Z &=&    \int \prod_i D\Phi_i \prod_j D\Psi_j
                e^{-\int d^2zd^2\theta W(\Phi_i,\Psi_j)+c.c.}\\
  &\sim& \int \prod D\xi_i
                e^{-\int d^2zd^2\theta W(\xi_i)+c.c.}
\eea
with superpotential
\beq
W=\xi_5(1+\xi_1\xi_6^3+\xi_2\xi_7^3) + \xi_1(1+\xi_2^3+\xi_3^3+\xi_4^3).
\eeq
Integrating out $\xi_1,\xi_5$ then leads to a product of two delta functions
whose arguments are interpreted as the two polynomials $p_1,p_2$ in
inhomogeneous coordinates.

A similar analysis can be performed for the manifolds (\ref{myclass}) of ref.
\cite{rs89}: starting from
\begin{equation}
W(\Phi_i,\Psi_i)=\Phi_1\Psi_1^2+\Phi_2\Psi_2^2+\sum_{i=1}^5 \Phi_i^{l/l_i},
\end{equation}
one can assume that $\Phi_1\neq 0$ and define the new coordinates
\begin{equation}
\xi_1= \Phi_1^{l_1},
{}~~~~ \xi_i=\frac{\Phi_i}{\Phi_1^{l_i/l_1}}, ~i=2,\cdots ,5,
{}~~~~ \xi_6=\Phi_1\Psi_1^2,
{}~~~~ \xi_7=\frac{\Psi_2}{\Psi_1}.
\end{equation}
This can be rewritten as
\begin{equation}
W(\xi_i)= \xi_6\left (1+\xi_2\xi_7^2\right ) +
          \xi_1\left (1 + \sum_{i=2}^5 \xi_i^{l/l_i}\right )
\end{equation}
and because the Jacobian is constant the path integral has not changed.
By integrating out the coordinates $\xi_1, \xi_6$ one finally obtains
the polynomial constraints (written in homogeneous coordinates)
\begin{eqnarray}
p_1 &=& y_1z_1^2 + y_2z_2^2, \nn  \\
p_2 &=& \sum_{i=1}^5 y_i^{l/l_i}.
\end{eqnarray}

Interpreting the coordinates $z_i$ as the variables of the projective curve
and the coordinates $y_i$ as the variables of a projective 4--fold these
constraints define a complete intersection in
\beq
\matrix{\IP_{(l_1,l_2,l_3,l_4,l_5)}\cr \IP_{(1,1)}\hfill\cr}
\left[\matrix{l_1&l\cr 2&0\cr}\right]
\eeq
where $l=l_2+l_3+l_4+l_5$.

This is not the whole story however, since we have not yet computed
the spectrum
of the theory. By correcting the Euler numbers of these configurations
\begin{equation}
\chi_s =-\frac{2l}{l_1\cdots l_5}\left [l_1^3
           + \sum_{i=2}^5 l_i^2 {\hat l}_i
           + \sum_{i<j<k \atop i,j,k\in \{2,..,5\}} 2l_il_jl_k \right ]
\end{equation}
where ${\hat l}_i=l-l_i$, with the contributions coming from the blow--up
of the singular sets one finds agreement with the Landau--Ginzburg result
as described in \cite{rs89}.  It is furthermore possible to find the
exactly solvable models for these manifolds as well:
to each of these manifolds one can associate an exactly solvable model by
starting from the weights $l_i$ of the weighted projective space.
Defining the levels as
\begin{equation}
k_1=k_2 = \frac{2l}{l_1}-2,~~k_i=\frac{l}{l_i}-2,~~i=3,4,5
\end{equation}
one finds for the central charge of these models
\begin{equation}
c= 2\frac{3k_1}{k_1+2} + \sum_{i=3}^5 \frac{3k_i}{k_i+2} = 9
\end{equation}
as is necessary for an allowed configuration. Finally one may check the
spectra of the various incarnations of the individual models.

It is clear from the discussion in Section 3 that via the analysis of the
singular sets of the variety, a prescription emerges for the decomposition of
the superpotential $W$ into the $Q$ polynomials which eventually define the
Calabi--Yau manifold of critical dimension.
We may consider the singular sets either in the higher dimensional weighted
hypersurface  or in the corresponding affine variety, thereby translating the
analysis appropriately. For the manifold (\ref{smf}) we are then lead to
the affine hypersurface
\beq
\IC_{(2,2,2,3,3,3,3)}[9]
\eeq
whose singular sets are
\beq
\IC_3, ~~~~ \IC_4[3].
\eeq
whereas for the manifolds of type (\ref{myclass}) we find
\beq
\IC_{(2k,K-k,2k,K-k,2k_3,2k_4,2k_5)}[2K]
\eeq
with (generic) singular sets
\beq
\IC_1,~~~~\IC_{(k,k,k_3,k_4,k_5)}[K].
\eeq
The details of the construction of \cite{gvw89} remain in place, of course,
but what emerges is the explanation that the path integral in fact localizes
on the singular sets of the affine space over which the superpotential lives.

\vskip .2truein
\noindent
\section{Mirrors of Rigid String Vacua}

\noindent
In the framework of the Landau--Ginzburg theoretic description of string vacua
a number of models appear whose spectrum makes them candidates for mirrors of
rigid ground states.
Indeed, by applying the construction of \cite{ls90} it is easy to find
the corresponding pairs of Landau--Ginzburg potentials which define the
appropriate  mirror partners of rigid theories, as will be seen in this
Section.

In the list of Landau--Ginzburg theories with isolated singularities
\cite{ks94, krsk92} there appear two distinct models without antigenerations,
both of which were mentioned in Section 3
\beq
(1^9)_{A^9}^{(0,84)} \leftrightarrow \IC^*_{(1,1,1,1,1,1,1,1,1)}[3]
 \ni  \left\{ W=\sum_{i=1}^3 z_i^3 =0 \right\}
\eeq
and
\beq
(2^6)_{A^6}^{(0,90)} \leftrightarrow  \IC^*_{(1,1,1,1,1,1,2)}[4]
 \ni  \left\{W=\sum_i z_i^4 + z_7^2 =0\right\}.
\eeq

The mirrors of these theories can be constructed by using the results of
\cite{ls90}. Consider e.g. the orbifold
\beq
\IC_{(1,1,1,1,1,1,1,1,1)}[3]/\ZZ_3^6:
{\scriptsize \left[\matrix{2&1&0&0&0&0&0&0&0\cr
                                     0&2&1&0&0&0&0&0&0\cr
                                     0&0&2&1&0&0&0&0&0\cr
                                     0&0&0&2&1&0&0&0&0\cr
                                     0&0&0&0&2&1&0&0&0\cr
                                     0&0&0&0&0&2&1&0&0\cr}\right].
}
\eeq
of the model $\IC_{(1,1,1,1,1,1,1,1,1)}[3]$. This orbifold has
the spectrum $(h^{(1,1)},h^{(2,1)},\chi)=(84,0,168)$ and leads, via
fractional transformations
\cite{ls90} of the coordinates, to the Landau--Ginzburg potential
\beq
\IC_{(96,48,72,60,66,63,43,64,64)}[192] \ni
\left\{z_1^2+z_1z_2^2+z_2z_3^2+z_3z_4^2+z_4z_5^2+z_5z_6^2+z_6z_7^3+z_8^3+
       z_9^3=0\right\}
\eeq
which has the correct spectrum $(84,0,168)$.

Even though the above mirror is certainly a valid representation it is
not a unique one. Consider the orbifold
\beq
\IC_{(1,1,1,1,1,1,1,1,1)}[3]/\ZZ_3^7:
{\footnotesize \left[\matrix{2&1&0&0&0&0&0&0&0\cr
                                       0&2&1&0&0&0&0&0&0\cr
                                       0&0&2&1&0&0&0&0&0\cr
                                       0&0&0&2&1&0&0&0&0\cr
                                       0&0&0&0&2&1&0&0&0\cr
                                       0&0&0&0&0&2&1&0&0\cr
                                       0&0&0&0&0&0&2&1&0\cr}\right].
}\eeq
which also has the spectrum $(84,0,168)$. Applying fractional
transformations again shows that this orbifold is isomorphic to the
hypersurface
\beq
\IC_{(192,96,144,120,132,126,129,85,128)}[384] \ni
\left\{z_1^2+z_1z_2^2+z_2z_3^2+z_3z_4^2+z_4z_5^2+z_5z_6^2+z_6z_7^2+
       z_7z_8^3+ z_9^3=0\right\}.
\eeq

For the second theory the orbifolding
\beq
\IC^*_{(1,1,1,1,1,1,2)}[4]
{\large /} \ZZ_4^3:
{\footnotesize \left[\matrix{3&1&0&0&0&0&0\cr
                           0&3&1&0&0&0&0\cr
                           0&0&3&1&0&0&0\cr
                   0&0&0&3&1&0&0\cr}\right]
}
\eeq
leads to the mirror spectrum and fractional transformations of the
variables lead to the mirror potential
\beq
\IC_{(108,72,84,80,61,81,162)}[324] \ni
\{ z_1^3 + z_1z_2^3+z_2z_3^3+z_3z_4^3 + z_4z_5^4+z_6^4+z_7^2=0\}.
\eeq

All representations of the rigid vacua and their mirrors that appear in
the list of all Landau--Ginzburg vacua can in fact be mapped into each
other via fractional
transformations. The precise relations are collected in Table 1.

\begin{center}
\begin{small}
\begin{tabular}{||l c c r  l||}
\hline
LG--Theory &Spectrum &Group &Action &~~~~~~~~Mirror LG--Theory \tabroom \\
\hline
$\IC_{(1,1,1,1,1,1,1,1,1)}[3]$   &(0,84)
  & $\ZZ_3^6$: & {\scriptsize $\left[\matrix{2&1&0&0&0&0&0&0&0\cr
                           0&2&1&0&0&0&0&0&0\cr
                           0&0&2&1&0&0&0&0&0\cr
                           0&0&0&2&1&0&0&0&0\cr
                           0&0&0&0&2&1&0&0&0\cr
                           0&0&0&0&0&2&1&0&0\cr}\right]$ }
& $\IC_{(43,48,60,63,64,64,66,72,98)}[192]$   \tabroom \\
                                 &
& $\ZZ_3^7$: &{\scriptsize $\left[\matrix{2&1&0&0&0&0&0&0&0\cr
                           0&2&1&0&0&0&0&0&0\cr
                           0&0&2&1&0&0&0&0&0\cr
                           0&0&0&2&1&0&0&0&0\cr
                           0&0&0&0&2&1&0&0&0\cr
                           0&0&0&0&0&2&1&0&0\cr
                           0&0&0&0&0&0&2&1&0\cr}\right]$ }
& $\IC_{(57,64,80,84,85,86,88,96,128)}[256]$ \tabroom \\
                                &
& $\ZZ_3^8$: & {\scriptsize $\left[\matrix{2&1&0&0&0&0&0&0&0\cr
                           0&2&1&0&0&0&0&0&0\cr
                           0&0&2&1&0&0&0&0&0\cr
                           0&0&0&2&1&0&0&0&0\cr
                           0&0&0&0&2&1&0&0&0\cr
                           0&0&0&0&0&2&1&0&0\cr
                           0&0&0&0&0&0&2&1&0\cr
                           0&0&0&0&0&0&0&2&1\cr}\right]$ }
& $\IC_{(85,96,120,126,128,129,132,144,192)}[384]$ \tabroom \\
$\IC_{(1,1,1,1,1,1,2)}[4]$    &(0,90)
 & $\ZZ_4^4$: &  {\scriptsize $\left[\matrix{3&1&0&0&0&0&0\cr
                           0&3&1&0&0&0&0\cr
                           0&0&3&1&0&0&0\cr
                           0&0&0&3&1&0&0\cr}\right]$ }
 & $\IC_{(61,72,80,81,84,108,162)}[324]$ \tabroom \\
                              &
 & $\ZZ_4^4\times \ZZ_2$: & {\scriptsize $\left[\matrix{3&1&0&0&0&0&0\cr
                           0&3&1&0&0&0&0\cr
                           0&0&3&1&0&0&0\cr
                           0&0&0&3&1&0&0\cr
                           0&0&0&0&1&0&1\cr}\right]$ }
 &$\IC_{(54,60,61,63,76,81,91)}[243]$   \tabroom \\
                              &
 &$\ZZ_4^5$: & {\scriptsize $\left[\matrix{3&1&0&0&0&0&0\cr
                           0&3&1&0&0&0&0\cr
                           0&0&3&1&0&0&0\cr
                           0&0&0&3&1&0&0\cr
                           0&0&0&0&3&1&0\cr}\right]$ }
 & $\IC_{(91,108,120,122,126,162,243)}[486]$ \tabroom \\
                              &
 &$\ZZ_4^5\times \ZZ_2$: & {\scriptsize $\left[\matrix{3&1&0&0&0&0&0\cr
                           0&3&1&0&0&0&0\cr
                           0&0&3&1&0&0&0\cr
                           0&0&0&3&1&0&0\cr
                           0&0&0&0&3&1&0\cr
                           0&0&0&0&0&1&1\cr}\right]$ }
 & $\IC_{(72,80,81,84,101,108,122)}[324]$   \tabroom \\
\hline
\end{tabular}
\end{small}
\end{center}

\noindent
{\bf Table 1:} {\it Rigid Landau--Ginzburg Theories
      \fnote{4}{This table contains the distinct theories. The other
               theories in the list of all Landau--Ginzburg vacua, listed
               here are isomorphic to the ones in Table 1.
        \begin{small}
          \begin{center}
           \begin{tabular}{||l c c r l ||}
           \hline
    $\IC_{(1,1,1,2,2,2,3)}[6]$  &(0,84)
    &$\ZZ_6^2\times \ZZ_3$
    &{\scriptsize $\left[\matrix{5&1&0&0&0&0&0\cr
                                         0&5&1&0&0&0&0\cr
                                         0&0&2&1&0&0&0\cr}\right]$}
  &$\IC_{(48,60,63,79,100,100,150)}[300]$  \tabroom \\
                                      &
     &$\ZZ_6^2\times \ZZ_3$
    &{\scriptsize $\left[\matrix{5&1&0&0&0&0&0\cr
                                         0&5&1&0&0&0&0\cr
                                         0&0&2&1&0&0&0\cr
                                         0&0&0&2&1&0&0\cr}\right]$}
  &$\IC_{(96,120,121,126,200,237,300)}[600]$ \tabroom \\
                                &
  &$\ZZ_6^2\times \ZZ_3^3$
  &{\scriptsize $\left[\matrix{5&1&0&0&0&0&0\cr
                                         0&5&1&0&0&0&0\cr
                                         0&0&2&1&0&0&0\cr
                                         0&0&0&2&1&0&0\cr
                                         0&0&0&0&2&1&0\cr}\right]$}
   &$\IC_{(64,80,84,93,121,158,200)}[400]$   \tabroom \\
                                     &
  &$\ZZ_6\times \ZZ_2 \times \ZZ_3^3$
  &{\scriptsize $\left[\matrix{5&1&0&0&0&0&0\cr
                                       0&1&0&0&0&0&1\cr
                                       0&0&2&1&0&0&0\cr
                                       0&0&0&2&1&0&0\cr
                                       0&0&0&0&2&1&0\cr}\right]$}
  & $\IC_{(48,55,60,64,75,88,90)}[240]$  \tabroom \\
 \hline
\end{tabular}
\end{center}
\end{small}
 }}

Thus we have matched all the rigid vacua that appear in the construction
of all Landau--Ginzburg theories \cite{ks94}\cite{krsk92}.
Aside from the rigid vacua that arise in the Landau--Ginzburg formulation
of string vacua there exist of course also rigid vacua that are obtained
by orbifolding Landau--Ginzburg theories with respect to discrete groups.
More importantly however there exist further rigid Calabi--Yau manifolds
whose mirrors cannot be described in the Landau--Ginzburg framework.
Examples are furnished by the resolution of particular nontransverse
quintics constructed by Schoen \cite{s85} and van Straten \cite{vs93}.

Fractional transformations therefore suggest a detailed analysis of the
special Fano varieties that are associated to the Landau--Ginzburg orbifolds
in the table. More precisely it would be of interest to understand in detail
the following pairings:

\begin{center}
\begin{small}
\begin{tabular}{|| l l||}
\hline
\hline
Special Fano Varieties   &~~~~~~ Mirror Special Fano Varieties \tabroom \\
\hline
\hline
$\IP_8[3]$    &~~~~~~ $\IP_{(43,48,60,63,64,64,66,72,98)}[192]$ \tabroom \\
            &~~~~~~ $\IP_{(57,64,80,84,85,86,88,96,128)}[256]$ \\
            &~~~~~~ $\IP_{(85,96,120,126,128,129,132,144,192)}[384]$ \\
\hline
$\IP_{(1,1,1,1,1,1,2)}[4]$ &~~~~~~ $\IP_{(61,72,80,81,84,108,162)}[324]$
\tabroom \\
                          &~~~~~~ $\IP_{(54,60,61,63,76,81,91)}[243]$  \\
                          &~~~~~~ $\IP_{(91,108,120,122,126,162,243)}[486]$ \\
                          &~~~~~~ $\IP_{(72,80,81,84,101,108,122)}[324]$ \\
\hline
\hline
\end{tabular}
\end{small}
\end{center}

\centerline{{\bf Table 2:} {\it Rigid mirror pairings in the framework of
special
              Fano folds.}}

\noindent
\section{ Toric Considerations}

\noindent
The simplest of all Calabi--Yau spaces surely is the set
\beq
\IP_1[2]=\{z_1^2+z_2^2=0\}
\eeq
which happens to consist of two points exactly. The higher dimensional
special Fano manifold of this space is $\IP_3[2]$ whose toric analysis
is what follows. This clearly is just a toy example, but it is a toy example
of some consequence.

Writing
\bea
 p &=&x_1^2+\cdots +x_4^2 +x_1x_2 +\cdots x_3x_4 \nn \\
   &=& z_3z_4 \left(\frac{z_1^2}{z_3z_4} + \frac{z_2^2}{z_3}+
                    \frac{z_3}{z_4} + \frac{z_4}{z_3} +\frac{z_1z_2}{z_3z_4}
                   +\frac{z_1}{z_4}
                   + \frac{z_1}{z_3} + \frac{z_2}{z_3} + 1\right)
\eea
leads (in $U_4=\{x_4=1\}$) to the integral points
\beq
(2,0,-1),(0,2,-1),(0,0,1),(0,0,-1),(1,1,-1),(1,0,0),(1,0,-1),(0,1,0),
(0,1,-1),(0,0,0)
\eeq
which span a polyhedron whose vertices are
\beq
\Delta = \{(2,0,-1),(0,2,-1),(0,0,1),(0,0,-1)\}.
\eeq
This polyhedron has no integral point in its interior and hence one deduces
from it that $h^{(2,0)}=0$ for the quadratic surface.  This is good
(and something we knew because the surface has positive first Chern class.)
To obtain
\beq
h^{(1,1)}(\bZ)= l^*(2\Delta) - 4l^*(\Delta)-3-
       \sum_{codim~ \G=1} \left(l^*(\G)-1\right)
\eeq
the first task is to count the number of integral points in the
interior of the polyhedron $2\Delta$ which is spanned by
\beq
2\Delta = \{(4,0,-2),(0,4,-2),(0,0,2),(0,0,-2)\}
\eeq
After some computing one finds  that there is only one interior
integral point $(1,1,-1)$ and hence
\beq
l^*(2\Delta)=1.
\eeq
There are 4 two--dimensional faces with no interior integral points
and therefore
\beq
h^{(1,1)}(\bZ)= 1-3-4(-1)=2
\eeq
which we also knew because $\IP_3[2]=\IP_1\times \IP_1$.

The upshot of this toric analysis of the noncritical manifolds suggests that
the special Fano manifolds defined in (\ref{spefam}) lend themselves to
a toric description in terms of polyhedra $\Delta$  which are characterized by
the condition that the $Q$--fold has precisely one interior integral point
\beq
l^*(Q\Delta) =1.
\eeq

A much more detailed translation of our framework into the language of toric
geometry has been provided by Batyrev and Borisov \cite{bb94}.

\vskip .3truein
\section{Phases of Special Fano Manifolds}

\noindent
It clearly is of interest to find a $\si$--model theoretic framework
for our construction. In \cite{w93} an analysis of gauged N$=$2
supersymmetric theories has been presented which also applies to models
that do not correspond to Calabi--Yau spaces. This analysis pertains to
models which may be closely related to special Fano  manifolds.
In the following attention is restricted to smooth manifolds; these contain,
as we have seen, manifolds describing mirrors of rigid Calabi--Yau manifolds
and therefore have been the most puzzling.

The starting point of the analysis in \cite{w93} is a U(1) gauge theory in
N$=$2 superspace, extending the standard Landau--Ginzburg action for the
chiral N$=$2 superfields to
\beq
L=L_{kin,V} +  L_{kin,\Phi_i} + L_{W,\Phi_i} + L_{D,\theta}.
\eeq
Consider, in the notation of \cite{w93}, the gauge invariant field strength
\beq
\Si =\frac{1}{2\sqrt{2}} \{\bcD_+,\cD_-\}
\eeq
the kinetic term of which is given by
\beq
\cA_{g} = -\frac{1}{4e^2} \int d^2z d^4\theta~ \bSi \Si.
\eeq
There are two possible interactions, the $\theta$ angle and the
Fayet--Illiopoulos D--term. These can be written as
\beq
\cA_{{\rm D},\theta} = \frac{it}{2\sqrt{2}} \int d^2z d\theta^+d\bth^-~ \Si
               + h.c.
\eeq
where
\beq
t=ir + \frac{\theta}{2\pi}
\eeq
where $r$ is the coefficient of the D--term.

To this are added $N$ chiral superfields with U(1)--charge $Q_i$.
The kinetic energy of these fields is chosen to be
\beq
\cA_{m} = \int d^2z d^4\theta ~ \sum_i \Phi_i \Phi_i
\eeq
and the superpotential is assumed to be of gauge invariant form
\beq
\cA_W = -\int d^2z d^2\theta ~W(\Phi_i) - h.c.
\eeq
which is supersymmetric because the $\Phi_i$ are chiral and $W$ is
holomorphic.

The constant part of the lowest components of the superfields $\Phi_i$
will be thought of as parametrizing the $n$--dimensional complex space
$\IC_n$, assuming, as is done in
\cite{w93}, that  the K\"ahler metric in the kinetic term of the $\Phi_i$
should be flat.

The bosonic equations of motion for the auxiliary fields $D_a$ and
$F_i$ become
\beq
D = -e^2 \left(\sum_i Q_i |\phi_i|^2 -r\right)
\eeq
and
\beq
F_i = \frac{\del W}{\del \phi_i}.
\eeq

The bosonic potential that one obtains in terms of the matter fields
$\phi_i$ and the auxiliary fields $D_a$ and $F_i$ is
\beq
U(\phi_i,\si)=\frac{1}{2e^2} D^2
              + \sum_i \left|\frac{\del W}{\del \phi_i}\right|^2
               + 2|\si|^2 \sum_i Q_i^2 |\phi_i|^2.
\eeq
Including the one--loop correction of the theory leads to a
$\si$--dependent effective $r$
\beq
r_{\rm eff} \sim r +\frac{\sum_i Q_i}{2\pi} ln\left(\frac{|\si|}{\mu}\right).
\eeq

Assume now that the superpotential takes the form\beq
W(\Phi_i) = \Phi_0 \tW(\Phi_1,....,\Phi_N)
\eeq
where $\tW$ is a quasihomogeneous polynomial which is assumed to be
transverse, i.e. the equations
\beq
\frac{\del \tW}{\del \Phi_i} =0
\eeq
can only be solved at the origin.

With this potential the bosonic potential becomes
\beq
U(\phi_i,\si)=\frac{1}{2e^2} D^2 + \left|\tW\right|^2
             +|\phi_0|^2  \sum_i \left|\frac{\del \tW}{\del \phi_i}\right|^2
               +2|\si|^2\left(\sum_ik_i^2|\phi_i|^2
                              +k_0^2 |\phi_0|^2 \right)
\lleq{bospot}
with
\beq
D = -e^2 \left(\sum_i k_i |\phi_i|^2 -k_0 \phi_0 \bphi_0 - r_{\rm eff} \right).
\eeq

Note that all terms in (\ref{bospot}) are $\geq 0$. Thus in order to minimize
the potential $U$ one has to minimize $D^2$ which means different things,
depending on what the variable $r$ does.  Assume now, that all chiral
superfields $\phi_i$ have charge unity under the U(1) gauge group.

\noindent
The $r_{\rm eff}=r>>0$ phase:
In this case not all $\phi_i$ can be zero. But since the polynomial $p$ is
transverse everywhere except at the origin this means, because $\del \tW$
is nonzero for some $i$, that $\phi_0$ must be zero. Thus $D=0$ leads to
 \beq
\sum_{i=1}^N \bphi_i \phi_i = r.
\eeq
This simply defines a sphere $S^{2N-1} \subset \IC_N$.
Recalling that one has to mod out the U(1) gauge group and also that the
sphere $S^{2N-1}$ can be Hopf--fibered $S^1 \lra S^{2N-1} \lra \IP_{N-1}$
leads to the condition that the constant bosonic components of $\Phi_i$
parametrize a projective space $\IP_{N-1} = S^{2N-1}/U(1)$.
Furthermore the vanishing of $\tW$ leads to a geometry for the space of
ground states described by a hypersurface embedded in $\IP_{N-1}$.

\noindent
The $r_{\rm eff}= r<<0$ phase:
In this case the vanishing of $D$ leads to $\phi_0\neq 0$
and hence the term $|\phi_0|^2 \sum_i |\del_i \tW|^2$ enforces that $\phi_i=0$
since this is the only place where the partials are allowed to vanish, because
of transversality. This fixes the modulus of $\phi_0$ to be
\beq
|\phi_0| = \sqrt{-r}{N}.
\eeq
Because of the gauge invariance the classical vacuum is in fact unique, modulo
these gauge transformations. Expanding around this vacuum leads to
massless $\phi_i$
(for $N\geq 3$). To find the potential for these massless fields one has to
integrate out the massive field $\phi_0$. Integrating out $\phi_0$ means
setting $\phi_0$ to its expectation value. Thus the effective
superpotential of the low energy theory is
\beq
\tW = \sqrt{-r} W(\phi_i).
\eeq
The factor $\sqrt{-r}$ is inessential since it can be absorbed by rescaling the
$\phi_i$. Since the origin is a multicritical point, this describes a
Landau--Ginzburg theory. More precisely it describes a Landau--Ginzburg
orbifold.

Applying this analysis to the two smooth higher dimensional manifold,
$\IP_8[3]$, with Hodge diamond
{\scriptsize
\begin{center}
\begin{tabular}{c c c c c c c c c c c c c c c}
  &  &  &  &   &  &    &1  &    &   &   &   &   &   &   \nn \\
  &  &  &   &   &  &0   &   &0   &   &   &   &   &   &   \nn \\
  &  &  &   &   &0 &    &1  &    &0  &   &   &   &   &   \nn \\
  &  &  &   &0  &  &0   &   &0   &   &0  &   &   &   &   \nn \\
  &  &  &0  &   &0 &    &1  &    &0  &   &0  &   &   &   \nn \\
  &  &0 &   &0  &  &0   &   &0   &   &0  &   &0  &   &   \nn \\
  &0 &  &0  &   &0 &    &1  &    &0  &   &0  &   &0  &    \nn \\
0 &  &0 &   &1  &  &84  &   &84  &   &1  &   &0  &   &0  \nn \\
  &0 &  &0  &   &0 &    &1  &    &0  &   &0  &   &0  &    \nn \\
  &  &0 &   &0  &  &0   &   &0   &   &0  &   &0  &   &   \nn \\
  &  &  &0  &   &0 &    &1  &    &0  &   &0  &   &   &   \nn \\
  &  &  &   &0  &  &0   &   &0   &   &0  &   &   &   &   \nn \\
  &  &  &   &   &0 &    &1  &    &0  &   &   &   &   &   \nn \\
  &  &  &   &   &  &0   &   &0   &   &   &   &   &   &   \nn \\
  &  &  &   &   &  &    &1  &    &   &   &   &   &   &   \nn \\
\end{tabular},
\end{center}
}
\noindent
and  $\IP_{(1,1,1,1,1,1,2)}[4]$, with Hodge diamond
{\scriptsize
\bea
\begin{tabular}{c c c c c c c c c c c c c c c}
  &  &  &   &   &  &    &1  &    &   &   &   &   &   &   \nn \\
  &  &  &   &   &  &0   &   &0   &   &   &   &   &   &   \nn \\
  &  &  &   &   &0 &    &1  &    &0  &   &   &   &   &   \nn \\
  &  &  &   &0  &  &0   &   &0   &   &0  &   &   &   &   \nn \\
  &  &  &0  &   &0 &    &1  &    &0  &   &0  &   &   &   \nn \\
  &  &0 &   &1  &  &90   &   &90  &   &1  &   &0  &   &   \nn \\
  &  &  &0  &   &0 &    &1  &    &0  &   &0  &   &   &   \nn \\
  &  &  &   &0  &  &0   &   &0   &   &0  &   &   &   &   \nn \\
  &  &  &   &   &0 &    &1  &    &0  &   &   &   &   &   \nn \\
  &  &  &   &   &  &0   &   &0   &   &   &   &   &   &   \nn \\
  &  &  &   &   &  &    &1  &    &   &   &   &   &   &   \nn \\
\end{tabular},
\eea
}

\noindent
leads to the conclusion that there are two, respectively one, additional
mode(s) in these theories that would not be present if we were to consider
the relevant string spectrum.
These numbers fit perfectly when considering the Hodge diamond
decomposition of the cohomology of the higher dimensional special Fano
manifold into a critical string part and a remaining part, as described in
the previous Sections.

The above counting generalizes:
for smooth special Fano manifolds of dimension $D_{crit}+2(Q-1)$ the
vertical cohomology contains (for $D_{crit}=3$)
\beq
H^{(1,1)},\dots, H^{(Q,Q)},
\eeq
leaving $(Q-1)$ forms unaccounted for.
Thus the counting of states described in \cite{w93} accounts precisely
for the additional cohomology groups found in the special class of Fano
folds (\ref{spefam}) which suggests that it is the proper
$\si$--model theoretic framework for these manifolds.

\vskip .3truein
\noindent
\section{Special Complete Intersection Fano Manifolds}

\noindent
The discussion in Section 3 of special Fano--folds which lead to
critical manifolds of codimension three and larger indicates that
the class of (\ref{spefam}) admits a generalization to noncritical
complete intersections.

More precisely, the analysis there suggests to consider the class
of manifolds
\beq
\matrix{\IP_{(k_1^1,\dots ,k_{n_1+1}^1)} \hfill \cr
        \IP_{(k_1^2,\dots ,k_{n_2+1}^2)} \hfill \cr
        \vdots \cr
        \IP_{(k_1^F,\dots ,k_{n_F+1}^F)} \hfill \cr
        \IP_{(k_1^{F+1},\dots ,k_{n_{F+1}+1}^{F+1})} \hfill \cr
        \vdots \cr
        \IP_{(k_1^{F+L},\dots ,k_{n_L+1}^{F+L})} \hfill \cr}
\left [
\matrix{d_1^1&d^1_2  &\ldots &d_f^1      &0             &\ldots &0\cr
        d_1^2&d_2^2  &\ldots &d_f^2      &0             &\ldots &0\cr
      \vdots &\vdots &\ddots &\vdots     &0             &\ldots &0\cr
        d_1^F&d_2^F  &\ldots &d_f^F      &0             &\ldots &0\cr
  0    &\ldots &0       &d_f^{F+1}  &d_{f+1}^{F+1} &\ldots &d_{f+K}^{F+1} \cr
      \vdots &\ddots &\vdots  &\vdots     &\vdots        &\ddots &\vdots \cr
  0    &\ldots &0       &d_f^{F+L}  &d_{f+1}^{F+L} &\ldots &d_{f+K}^{F+L} \cr
        }
\right]
\eeq
such that
\beq
c_1(M) = (Q-1-f) \sum_{i=f+1}^{f+K} c_1(\cN_i)
\eeq
for some integer $Q\in \IN$.

As an example consider the 7--fold
\beq
\IP_{(5,6,4,5,6,4,4,6,8)}[16]~\ni~
\left\{\sum_{i=1}^2 \left(x_i^2y_i + y_i^2z_i +z_i^4\right)
 +y_3^2z_3+z_3^4=0\right\},
\eeq
the $\ZZ_2$--fibering of which leads to the weighted complete
intersection
\beq
\matrix{ \IP_1\hfill \cr \IP_{(3,3,2,2,2,3,4)} \cr}
\left[\matrix{2&0\cr 3&8\cr}\right].
\eeq
This variety is not a Calabi--Yau space since its first Chern class
is nonvanishing
\beq
c_1 = 8h_2,
\eeq
where $h_2$ is the pullback of the K\"ahler class of the weighted
ambient factor. Furthermore, because $c_1(M)=c_1(N_2)$ it follows that $Q=3$
and the Calabi--Yau manifold is expected to be of codimension three.

Indeed, the $\ZZ_3$ fibering of this space leads to the space
\beq
\matrix{\IP_1 \hfill \cr \IP_2 \hfill \cr  \IP_{(1,1,1,2)} \cr}
\left[\matrix{2&0&0\cr 1&2&0\cr 0&1&4\cr}\right] \ni
\left\{ \begin{array}{c l}
        p_1=& \sum_{i=1}^2 x_i^2y_i=0 \\
        p_2=& \sum_{i=1}^3 y_iz_i =0 \\
        p_3=& \sum_{j=1}^3 z_i^4+z_4^2 =0
         \end{array}
\right\}
\eeq
which does define a Calabi--Yau 3--fold of codimension 3.

A further simple example in this class is the noncritical  manifold
\beq
\matrix{\IP_1\hfill \cr \IP_{(1,1,1,2,2,2)}\cr}
\left[\matrix{2 &0\cr 1&4\cr}\right]
\eeq
with first Chern class
\beq
c_1 = 4h_2.
\eeq
Applying the paradigm leads to the critical complete intersection
\beq
\matrix{\IP_1 \cr \IP_2 \cr \IP_2\cr}
\left[\matrix{2&0&0\cr 1&2&0\cr 0&1&2\cr}\right]
\eeq
which describes a K3 surface.

\vskip .2truein
\noindent
\section{Conclusion}

\noindent
It has been clear for some time now that, because of the existence of rigid
(2,2)--string vacua  mirror symmetry cannot be understood in the framework of
Calabi--Yau manifolds. Instead there must exist, beyond the realm of such
spaces,
a new type of variety which does contain information about critical vacua,
such as the spectrum of the massless fields, for both, Calabi--Yau type
ground states as well as for vacua which describe mirrors of rigid
Calabi--Yau manifolds.

It has been shown in \cite{rs93} and the present paper
that the special class of Fano manifolds of type (\ref{spefam})
generalizes the framework of Calabi--Yau vacua in the desired way: For
particular types of such Fano manifolds Calabi--Yau spaces of
critical dimension are embedded algebraically in a fibered submanifold.
For string vacua which cannot be described by
K\"ahler manifolds and which are mirror candidates of rigid Calabi--Yau
manifolds the higher dimensional manifolds still lead to the
spectrum of the critical vacuum and a rationale emerges that explains why a
Calabi--Yau representation is not possible in such theories.
Thus these manifolds of dimension $c/3 +2(Q-1)$ define an appropriate
framework in which to discuss mirror symmetry.

One intriguing  consequence that follows from our results suggests the
generalization of a conjecture regarding the relation between (2,2)
superconformal field theories of central charge $c=3D$, $D\in \IN$,
with N=1 spacetime supersymmetry on the one hand and K\"ahler manifolds
of complex dimension $D$ with vanishing first Chern class on the other.
It was suggested by Gepner \cite{g87a} that this relation is 1--1.
Our analysis indicates that instead superconformal theories of such a
 type are in correspondence
with K\"ahler manifolds of dimension $c/3 +2(Q-1)$ with a first Chern
class quantized in multiples of the degree of the defining polynomial.

The results described here indicate that Fano manifolds of the special type
considered in this paper are  not just  auxiliary devices but may be as
physical as  Calabi--Yau manifolds of critical dimension.

\vskip .3truein
\noindent
\section*{Acknowledgement}

\noindent
Part of this work was done at the Institute for Theoretical Physics,
Santa Barbara, supported in part by NSF grant PHY--89--04035, the
Theory Group at the University of Texas, and the Aspen Institute for Physics.
I'm grateful to these institutions for support and hospitality and to
V.V. Batyrev, P. Berglund, P. Candelas, E. Derrick, M. Lynker,
A. Strominger and E. Witten for discussions.


\end{document}